\documentclass[]{aa}  

\usepackage[utf8]{inputenc}
\usepackage{graphicx}
\usepackage{txfonts}
\usepackage{siunitx}
\usepackage{hyperref}
\usepackage{multirow}
\usepackage[dvipsnames]{xcolor} 
\usepackage{soul}
\usepackage{mathtools}

\usepackage[normalem]{ulem}

\begin{document} 
   \title{Combining the CLAUDS \& HSC-SSP surveys}
   \subtitle{$U+grizy(+YJHK_s)$ photometry and photometric redshifts for 18M galaxies in the $20~{\rm deg}^2$ of the HSC-SSP Deep and ultraDeep fields}

\author{ G. Desprez\inst{1,2}\thanks{\email{guillaume.desprez@smu.ca}} 
        \and V. Picouet \inst{3,4}
        \and T. Moutard \inst{3}
        \and S. Arnouts \inst{3}
        \and M. Sawicki \inst{2}\thanks{Canada Research Chair}
        \and J. Coupon\inst{1} 
        \and S. Gwyn\inst{5}
        \and L. Chen\inst{2}
        \and J. Huang\inst{6}
        \and A. Golob \inst{2}
        \and H. Furusawa \inst{7}
        \and H. Ikeda \inst{8}
        \and S. Paltani \inst{1}
        \and C. Cheng\inst{6}
        \and W. Hartley \inst{1}
        \and B. C. Hsieh \inst{9}
        \and O. Ilbert \inst{3}
        \and O. B. Kauffmann \inst{3}
        \and H. J. McCracken \inst{10}
        \and M. Shuntov \inst{10}
        \and M. Tanaka \inst{7}
        \and S. Toft \inst{11,12}
        \and L. Tresse \inst{3}
        \and J. R. Weaver \inst{11,12}
}

\institute{Department of Astronomy, University of Geneva, ch. d'\'Ecogia 16, CH-1290 Versoix, Switzerland
        \and Institute for Computational Astrophysics and Department of Astronomy and Physics, Saint Mary's University, Halifax, Nova Scotia, B3H 3C3, Canada
        \and Aix Marseille Univ, CNRS, CNES, LAM, Marseille, France
        \and Department of Astronomy, Columbia University, 550 W. 120$^{th}$ Street, New York, NY 10027, USA
        \and{NRC-Herzberg, 5071 West Saanich Road, Victoria, British Columbia V9E 2E7, Canada}
        \and{Chinese Academy of Sciences South America Center for Astronomy, National Astronomical Observatories, CAS, Beijing 100101, People’s Republic of China}
        \and{National Astronomical Observatory of Japan, 2-21-1 Osawa, Mitaka, Tokyo 181-8588, Japan}
        \and{National Institute of Technology, Wakayama College, 77 Noshima, Nada-cho, Gobo, Wakayama 644-0023, Japan}
        \and{Institute of Astronomy \& Astrophysics, Academia Sinica, Taipei 10617, Taiwan} 
        \and{Institut d’Astrophysique de Paris, UMR 7095, CNRS, and Sorbonne Universite\'e, 98 bis boulevard Arago, F-75014 Paris, France}
        \and{Cosmic Dawn Center (DAWN)}
        \and{Niels Bohr Institute, University of Copenhagen, Jagtvej 128, 2200 Copenhagen, Denmark}}

   \date{Received date; accepted date}

 
 \abstract{
We present the combination of the Canada-France-Hawaii Telescope (CHFT) Large Area $U$-bands Deep Survey (CLAUDS) and the Hyper-Suprime-Cam (HSC) Subaru Strategic Program (HSC-SSP) data over their four deep fields. We provide photometric catalogs for  $u$, $u^*$ (CFHT--MegaCam), $g$, $r$, $i$, $z$, and $y$ (Subaru--HSC) bands over $\sim 20~{\rm deg}^2$, complemented in two fields by data from the Visible and Infrared Survey Telescope for Astronomy (VISTA) Deep Extragalactic Observations (VIDEO) survey and the UltraVISTA survey, thus extending the wavelength coverage toward near-infrared with VIRCAM $Y$, $J$, $H$, and $K_s$ observations over $5.5~{\rm deg}^2$. The extraction of the photometry was performed with two different softwares: the HSC pipeline \texttt{hscPipe} and the standard and robust \texttt{SExtractor} software. Photometric redshifts were computed with template-fitting methods using the new \texttt{Phosphoros} code for the \texttt{hscPipe} photometry and the well-known \texttt{Le Phare} code for the \texttt{SExtractor} photometry. The products of these methods were compared with each other in detail. We assessed their quality using the large spectroscopic sample available in those regions, together with photometry and photometric redshifts from COSMOS2020, the latest version of the Cosmic Evolution Survey catalogs. We find that both photometric data sets are in good agreement in $Ugrizy$ down to magnitude~$\sim26$, and to magnitude~$\sim24.5$ in the $YJHK_s$ bands. We achieve good performance for the photometric redshifts, reaching precisions of $\sigma_{NMAD} \lesssim 0.04$ down to ${m}_i\sim25$, even using only the CLAUDS and HSC bands. At the same magnitude limit, we measured an outlier fraction of $\eta \lesssim 10\%$ when using the $Ugrizy$ bands, and down to $\eta \lesssim 6\%$ when considering near-infrared data. The \texttt{hscPipe} plus \texttt{Phosphoros} pipeline performs slightly worse in terms of photometric-redshifts precision and outlier fraction than its \texttt{SExtractor} plus \texttt{Le Phare} counterpart, which has essentially been tracked down to differences in the photometry. Thus, this work is also a validation of the \texttt{Phosphoros} code. The photometric catalogs with the data and photometric redshifts from the two pipelines are presented and made publicly available.


   }
   \maketitle
%


\section{Introduction}
\label{sec:intro}

Deep, wide-area multiband imaging surveys play a pivotal role in our studies of the Universe and of the formation and evolution of its content (e.g., CANDELS, \citealt{Grogin2011}; COSMOS, \citealt{Scoville2007}; CFHTLS, \citealt{Hudelot2012}; DES, \citealt{Abbott2018}). They are expected to continue to do so for the foreseeable future given the large investment of resources into projects such as LSST on the Rubin Observatory \citep{Ivezic2019}, \textit{Euclid} \citep{Laureijs2011}, and Roman Space Telescope \citep{Akeson2019}. While spectroscopy — particularly when highly multiplexed — can yield detailed physical information on samples of distant galaxies, it is expensive in telescope time, which makes it difficult to assemble very large samples of faint objects. In contrast to spectroscopy, photometry is relatively inexpensive, extremely multiplexed when done with modern, large-area mosaic imagers, and is also significantly less affected by selection biases. 

Much can be learned from photometry, particularly when samples are large and contain many faint objects. In particular, photometry can be used to estimate redshifts of vast samples of distant galaxies (e.g., \citealt{Beck2016, Tanaka2018, Moutard2020}), to characterize key physical properties such as their stellar masses and star formation rates (SFRs; e.g., \citealt{Sawicki1998, Papovich2001, Laigle2016}), and even constrain star formation histories (e.g., \citealt{Pacifici2016, Iyer2019}). These measurements are key ingredients in studies of how galaxies form and evolve over cosmic time and motivate many of the modern photometric surveys.

The Hyper Suprime-Cam Strategic Survey Program (HSC-SSP, \citealt{Aihara2018a}), which recently completed observations on the Subaru telescope, represents the current state of the art in deep, wide-area imaging surveys. Of particular interest here are HSC-SSP’s Deep and ultraDeep layers, which cover 26 and 3.5 deg$^2$, respectively, in the $grizy$ filters to depths of $\sim$25--28 AB (signal to noise ratio S/N=5$\sigma$ in 2\arcsec\ apertures). This is an unprecedented combination of area and depth that allows studies of galaxy evolution as a function of redshift and environment that are largely insensitive to cosmic variance and that will be unsurpassed until several years into LSST’s science operations. 

However, the HSC-SSP $grizy$ photometry spans only the optical and lacks both the shorter and the longer wavelengths that are critical for a number of science applications. In particular, complementing $grizy$ imaging with $U$-band photometry provides better measurements of SFRs at all redshifts and significantly improves photometric redshift performance at $z<0.7$ and at $z\sim2-3$ because it allows the combined filter set to span the Balmer and Lyman breaks, respectively, at these redshifts ( see Fig.~9 and Sect.~4.2 of \citealt{Sawicki2019} for more discussion on the $U$-band benefits for photo-$z$s). The SFR and photometric redshift measurements in the HSC-SSP Deep and ultraDeep layers were indeed key motivators for the Canada-France-Hawaii Telescope (CFHT) Large Area U-band Deep Survey (CLAUDS; \citealt {Sawicki2019}), a major (68 dedicated nights plus archival data) imaging survey that complements, with $U$-band, the $grizy$ imaging in the Deep and ultraDeep layers of HSC-SSP with an overlap of $\sim20~{\rm deg}^2$ between both surveys. Meanwhile, at longer wavelengths, complementing the CLAUDS+HSC-SSP U+grizy photometry with near-infrared data would make measurements of galaxy stellar masses beyond $z\sim 1$ possible and further improve photometric redshift performance by adding coverage of the Balmer break at higher redshifts.

These ingredients --- uniform photometry, photometric redshifts, and galaxy physical parameters --- are key elements needed for many studies of distant galaxy populations and their evolution. For this reason, in this paper, we present our latest multiband catalogs that combine CLAUDS, HSC-SSP, and — where available — VIDEO (VISTA Deep Extragalactic Observations) \citep{Jarvis2013} and UltraVISTA \citep{McCracken2012} data in the HSC-SSP Deep and ultraDeep fields. In previous papers \citep{Sawicki2019, Moutard2020}, we presented CLAUDS+HSC-SSP catalogs for these fields based on shallower HSC-SSP photometry and less sophisticated technical treatment. These earlier catalogs have already been used for several science applications: for example, \citet{Halevi2019} used this earlier CLAUDS+HSC-SSP photometry to characterize the properties of an active galaxy nucleus (AGN) containing low-mass galaxy; \citet{Moutard2020} and \citet{Harikane2022} constrained ultra-violet luminosity functions; and \citet{Iwata2022} used them to constrain the Lyman continuum escape fraction from AGNs. Additionally, the data were used for star-galaxy classification \citep{Golob2021}, selection of low-luminosity galaxies \citep{Cheng2021}, photometric redshift estimation \citep{Sawicki2019, Moutard2020, Huang2020}, and constraining galaxy-galaxy merger fraction evolution \citep{Thibert2021}.

These earlier CLAUDS+HSC-SSP catalogs already demonstrated the usefulness of a deep, $U$-band-enhanced wide-area photometric dataset in the HSC-SSP Deep and ultraDeep fields. However, because of their developmental nature, we did not make these earlier catalogs publicly available to the community. Moreover, the depth of $grizy$ imaging used in their creation was significantly shallower than what has become available as the HSC-SSP survey accumulated data over time. Finally, for simplicity, our earlier catalogs did not include any NIR information even though deep NIR imaging. 

The goal of the present paper is then to provide well-characterized catalogs that contain $U+grizy$ (+NIR, where available) photometry, photometric redshifts %
in the Deep and ultraDeep areas of the CLAUDS+HSC-SSP surveys. To build these catalogs, we use two sets of codes. We use recently developed photometry and photo-$z$ methods, namely a modified version of the HSC-SSP photometric pipeline (\texttt{hscPipe}, \citealt{Bosch2018}) for photometry extraction and \texttt{Phosphoros} (Paltani et al., in prep.; \citealt{Desprez2020}) for photo-$z$ computation. We also provide photometry and photometric redshifts measured with widely used and long-established codes, namely \texttt{SExtractor} \citep{Bertin1996} and \texttt{Le Phare} \citep{Arnouts2002,Ilbert2006}. The comparison of the results is allowing the validation of the different methods.%
With the present paper, we are making both catalogs publicly available for the community. This work shows the overall agreement between both catalogs in terms of photometry and photometric redshifts. The availability of the two catalogs allows the users to select the one most suitable for their own interests or to use them in tandem to assess the reliability of results as advocated by \citet{Weaver2021} for COSMOS2020.   %

Throughout the paper, magnitudes are provided in the AB system. CLAUDS $u$ and $u^*$ bands are referred to as $U$ bands when no distinction is made between them (see Sect.~\ref{sec:data-overview} of this paper or Sect.~2.1 of \citealt{Sawicki2019}). We also assume  the \citet{PlanckXIII2016} cosmology with  ${\rm H}_{0}=67.74~{\rm km~s^{-1}~Mpc^{-1}}$, $\Omega_M=0.3089$, and $\Omega_\Lambda=0.6911$.


\section{Data}
\label{sec:data}

\subsection{Overview of photometric data}
\label{sec:data-overview}

\begin{figure}
    \centering
    \includegraphics[width=\linewidth]{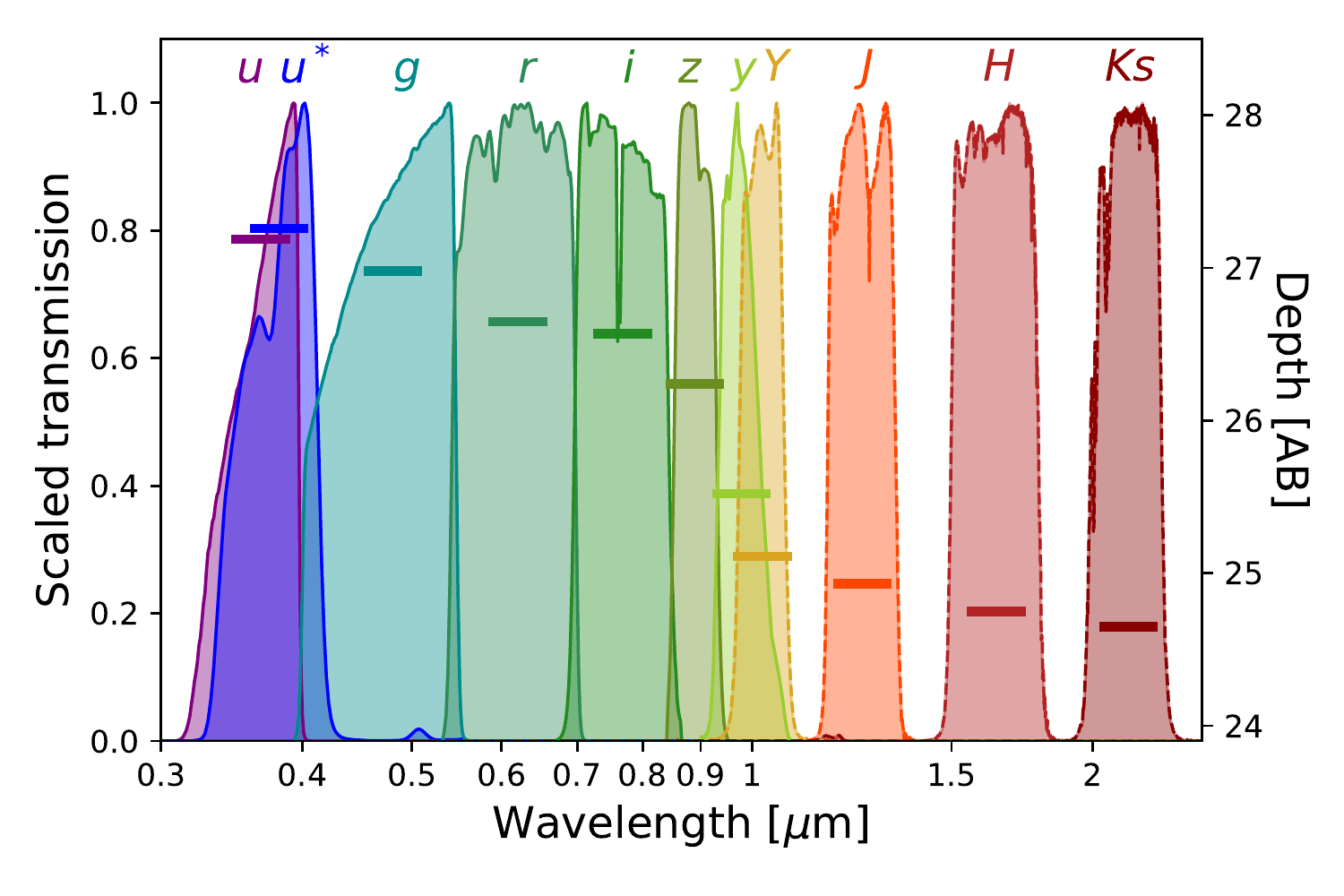}
    \caption{Transmission curves of the photometric bands accounting for instrument throughputs and typical atmospheric absorption. All curves are  scaled to 1. The horizontal lines indicate the 5$\sigma$ depths in 2\arcsec\ aperture measured for each band in the ultradeep regions for the E-COSMOS field. 
   }
    \label{fig:filter}
\end{figure}

\begin{figure*}
    \centering
    \includegraphics[width=\linewidth]{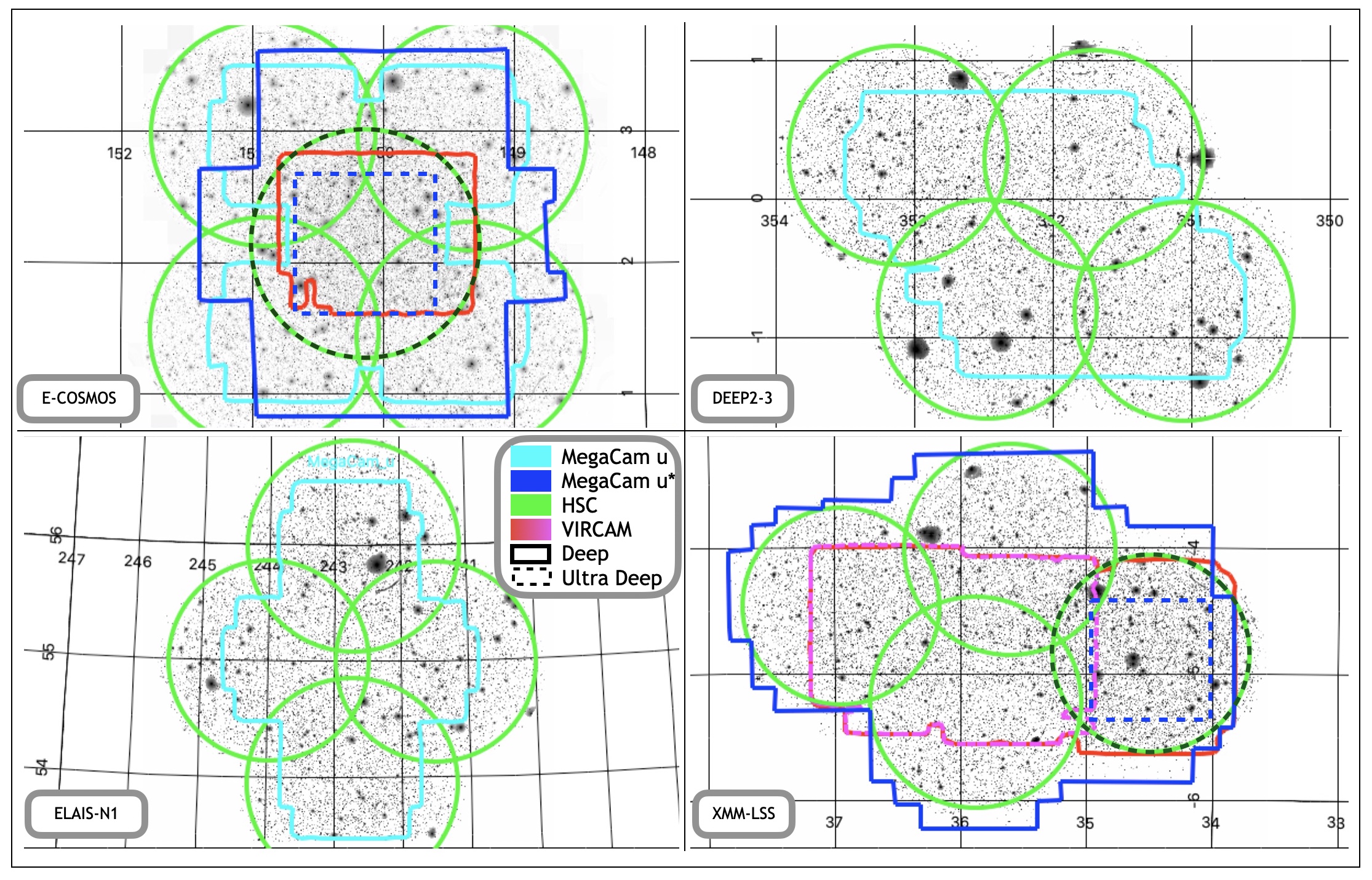}
    \caption{Detection images in the four CLAUDS-HSC fields. The footprints of the different observations are over-plotted in different colors: The CLAUDS Deep layer $u$ and $u^\star$ bands in blue; the HSC-SSP $grizy$ data in green; the VIDEO Near Infrared data in red (except in the XMM-LSS field where only two $J$ band pointings are available, colored in pink). The footprints of the  ultraDeep areas are shown with dashed lines.}
    \label{fig:layout}
\end{figure*}

The data consist of four fields (E-COSMOS, DEEP2-3, ELAIS-N1, and XMM-LSS; see Tables~\ref{tab:fields} and~\ref{tab:surveys}) and come from four different surveys. All the fields are covered by CLAUDS \citep{Sawicki2019} and the HSC-SSP \citep{Aihara2018a}, but E-COSMOS and XMM-LSS are also partly covered by two near-infrared (NIR) surveys, UltraVISTA \citep{McCracken2012} and VIDEO \citep{Jarvis2013}, respectively. The transmissions of the photometric bands of these data are shown in Fig.~\ref{fig:filter}, while their coverage and depth are presented in Fig.~\ref{fig:layout} and Table~\ref{tab:surveys}.

\begin{table}[]
    \centering
    \caption{Right ascension and declination coordinates (J2000) of the four deep field centers.}
    \begin{tabular}{l c c}
    \hline
    \hline
        \rule{0pt}{1.2em} & RA (J2000) & DEC(J2000) \\
    \hline
         \rule{0pt}{1.2em}E-COSMOS & $10:00:28.06$ & $+02:13:59.53$ \\
         DEEP2-3 & $23:28:25.18$ & $-00:11:55.21$ \\
         ELAIS-N1 & $16:11:17.05$ & $+55:02:40.85$ \\
         XMM-LSS & $02:22:28.96$ & $-04:44:24.20$ \\
    \hline
    \end{tabular}
    
    \label{tab:fields}
\end{table}

CLAUDS is a deep survey using the Canada-France-Hawaii-Telescope (CFHT) MegaCam imager \citep{Boulade2003} in the $u$ and $u^{*}$ bands.\footnote{The associated passband is simply noted $U$ when no distinction is made between the filters, following \citet{Sawicki2019}.} This program combines observing time from Canada, France, and China and consists of a total of 462 hours of observations, including dedicated CLAUDS observations plus archival $u^*$-band data in the COSMOS and XMM-LSS fields. The full survey covers a total area of 18.60~deg$^{2}$ over the four fields down to a median depth of $m_{U}=27.1$ (deep survey) 
 and two smaller but ultradeep areas covering regions of the E-COSMOS and XMM-LSS fields, reaching a median depth of $m_{U}=27.7$ over 1.36~deg$^{2}$. The ELAIS-N1 and DEEP2-3 fields are exclusively covered in the $u$ band, whereas the XMM-LSS is covered in the $u^{*}$ band. Only the central region of the E-COSMOS field has been observed in both $u^*$ and $u$ bands with a large overlap in the deep region, however only the $u^*$ observations have a distinct ultradeep region. All the details of the CLAUDS survey and data are presented in \citet{Sawicki2019}.

The HSC-SSP survey is a recently completed survey conducted with the Hyper Suprime-Cam camera \citep{Miyazaki2018} on the Subaru telescope in Hawaii. In this work, we consider the public data release 2 (PDR2) Deep and ultraDeep layers of the survey presented in \citet{Aihara2019}. The four fields have been observed in the $g$, $r$, $i$, $z$, and $y$ bands \citep{Kawanomoto2018} with median depths ranging from $m_g=27.3$ to $m_{y}=25.3$ for point sources detected at 5 $\sigma$ (all the depths are presented in \citealt[][Table 2]{Aihara2019}). 

The UltraVISTA survey is a NIR survey covering 1.5~deg$^2$ in the central region of the E-COSMOS field. We use the publicly available data release 3 images in the $Y$, $J$, $H$, and $K_s$ bands\footnote{\url{https://ultravista.org}}. The images were acquired by the VISTA Telescope \citep{Emerson2004} with the VIRCAM instrument \citep{Dalton2006}. The data reach depths of $m_{Y}\sim25$ and $m_{J, H, K_s}\sim24$ for $2\arcsec$-diameter apertures at 5 $\sigma$ \citep{McCracken2012}.

For the XMM-LSS field, we use the public VIDEO survey data release 4 images in the NIR $Y$, $J$, $H$, and $K_s$ bands\footnote{\url{http://www.eso.org/sci/observing/phase3/data_releases.html}}. As for UltraVISTA, the images were produced using the VIRCAM instrument on the VISTA Telescope. The VIDEO data cover around 4.5~deg$^{2}$, down to depths ranging from $m_Y=24.6$ to $m_{Ks}=23.8$ for 5~$\sigma$ in $2\arcsec$ apertures \citep{Jarvis2013}.

A summary of the data available in the different fields is presented in Table~\ref{tab:surveys}. The depths shown in the table are the medians of the $5\sigma$ scatter measured by us in 2\arcsec\ apertures thrown in the background of the images in all the patches (see Sect.\ref{sec:data-preparation} for patches definition). We note that our reported depths can differ from the ones announced by other teams, depending on the adopted definition. Indeed, the depths computed from point sources are deeper than those computed from the standard deviations of background fluxes in $2\arcsec$ apertures. 

\begin{table*}[]
    \centering
        \caption{Summary of the characteristics of the survey in the different fields: filter coverage, $5\sigma$, 2\arcsec\ depth and area. 
        }
    \begin{tabular}{l|cccccccc}
    \hline
    \hline
        \rule{0pt}{1.2em}Field & depth & complementary bands & area $U$ & depth $U$ & area HSC & depth HSC & area NIR &depth NIR \\
        & & & [deg$^2$] & $u$--$u^*$ & [deg$^2$] & $g$--$y$ & [deg$^2$] &$Y$--$K_s$\\
        \hline
         \rule{0pt}{1.2em}E-COSMOS & deep & $u$,$u^*$ & 4.6 & 27.2--26.3 & 4.8 & 26.4--24.7 & --- & --- \\
         & ultradeep & $u$,$u^*$,$Y$,$J$,$H$,$K_s$ & 0.8 &  27.2--27.3 & 1.7 & 27.0--25.5 & 1.4 &25.1--24.7 \\
         XMM-LSS & deep & $u^*$,$Y$,$J$,$H$,$K_s$ & 6.0 & 27.0 & 4.3 & 26.5--24.0 & 4.1 & 25.0--23.8 \\
         & ultradeep & $u^*$,$Y$,$J$,$H$,$K_s$ & 0.8 & 27.4 & 1.4 & 26.9--24.9 & --- & --- \\
         ELAIS-N1 & deep &  $u$ & 3.6 & 27.0 & 5.0 &26.5--24.5 & --- &---\\
         DEEP2-3 & deep & $u$ & 3.7 & 27.1 & 5.6 & 26.4--24.6 & --- & ---\\
         \hline
    \end{tabular}
    \label{tab:surveys}
\end{table*}

\subsection{Spectroscopic data}
\label{sec:data-spectro}
 The spectroscopic-redshift sample is a compilation of released  spectroscopic surveys, including:
 the SDSS-BOSS surveys \citep[Data Release 16,][]{SDSS_DR16};
 the GAMA survey \citep[Data release 3,][]{Baldry2018}; 
 the WiggleZ survey  \citep[final release, ][]{Drinkwater2018}; 
 the VVDS Wide and Deep surveys \citep{LeFevre2013}; 
 the VUDS  survey \citep{LeFevre2015}; 
 the DEEP2 survey \citep[Data Release 4,][]{Newman2013};
 the VIPERS survey \citep[Data Release 2,][]{Scodeggio2018};   
 the VANDELS survey \citep[Data Release 4,][]{Garilli2021};
 the CLAMATO survey \citep[Data Release 1,][]{KGLee2018};
 the UDSz survey \citep[][]{McLure2013,Bradshaw2013}.
 
We also include private spectroscopic observations from
the spectroscopic follow-up of faint low redshift galaxies (r$\le$22 and z$\le$0.5) in three of the HSC fields with the Hectospec spectrograph on the MMT telescope \citep{Cheng2021}; and from
the spectroscopic redshift catalog from the COSMOS team (M. Salvato, private communication), which consists of several optical and NIR spectroscopic follow-ups of X-ray to far-IR/radio sources, high-redshift star-forming and passive galaxies, as well as poorly represented galaxies in multidimensional color space \citep[C3R2,][]{Masters2019}.  We also include the low-resolution spectroscopic redshifts from slitless spectroscopy with the near-infrared grism from the 3DHST survey data release v4.1.5 using only those classified as secure grism redshift measurements  \citep{Momcheva2016,Skelton2014}.
  
For all the above redshift surveys, we compare our photometric redshifts with the most secure spectroscopic redshifts only, identified with high signal-to-noise and with several spectral features (i.e., equivalent to VVDS or VIPERS redshift flags 3 and 4).

\subsection{Data preparation}
\label{sec:data-preparation}

The PDR2 of HSC-SSP has been processed with version 6 of the \texttt{HSC/LSST pipeline} \citep{Bosch2018,Aihara2019}. 
We process the $U$ and NIR bands using the same version of the pipeline, which has been modified in order to process non-HSC data.%

The first step is to project the images onto the HSC reference pixel grid. As presented in \citet{Bosch2018},  the \texttt{hscPipe} divides the sky into \emph{tracts}, themselves divided in  \emph{patches}. A patch has a resolution of $4200 \times 4100$ pixels, with a pixel scale of $\ang{;;0.168}$/pixel. A tract is composed of $9 \times 9$ patches, all having a 100-pixel overlap with the adjacent patches and sharing the same tract coordinate system. The tracts form a mosaic on the sky, with some overlaps varying depending on the declination. The $U$-band data reduction uses the same HSC data format (i.e., tract patch and pixel grid) for the final products. The full details of U-band image processing are given in \citet{Sawicki2019}, but we review the key steps below. The detrended MegaCam images were processed using the \texttt{MEGAPIPE} data pipeline \citep{Gwyn2008} for astrometric and photometric calibration and for the stacking of the images. Two versions of the CLAUDS images are produced, one using the Pan-STARRS astrometry and another one using the more recent \textit{Gaia} astrometry \citep{Gaia2016a,Gaia2016b}. The processing done with the \texttt{hscPipe} described in Sect.~\ref{sec:photometry-hsc} uses the Pan-STARRS calibration for the U band and the processing done with \texttt{SExtractor} described in Sect.~\ref{sec:photometry-sextractor} uses the images calibrated with \texttt{Gaia}.
Individual images underwent photometric calibration before the production of stacks, generated directly on the HSC pixel grid. The images were resampled using \texttt{Swarp} \citep{Bertin2002}, the background subtracted using a $128\times128$ pixel mesh grid done with \texttt{Swarp}, and stacked to produce the images and weight maps matching the HSC patches.
In the case of the NIR data, the fully calibrated public mosaics (image and weight maps) are projected onto the HSC pixel grid with \texttt{Swarp}
for all the patches covered by the NIR data. For the VIDEO data, some tiles are duplicated due to the layout and overlap of the different observations. The duplicated tiles are combined by averaging them over the overlaps to obtain a single tile per patch.

 The $U$ and NIR band processed data are formatted into coadd files that can be handled by \texttt{hscPipe} (see \citealt{Bosch2018}).
 This is performed with a module developed for \texttt{hscPipe} \footnote{\url{https://github.com/jcoupon/importExtData}}.
 For each patch and band, the corresponding coadd file is gathering the images, and the variance map is computed from the weight map.  A mask of the bright stars found in the patch is created using the prescription of \citet{Coupon2018}. The point spread function (PSF) of the patch image is estimated. Finally, the coadd file is provided to the \texttt{hscPipe} for source detection and photometry extraction.
\subsection{Masking}
\label{sec:data-mask}

\subsubsection{Bright-star masks}
\label{sec:data-mask-brigh}

\begin{figure*}
     \centering
     \includegraphics[width=1\linewidth]{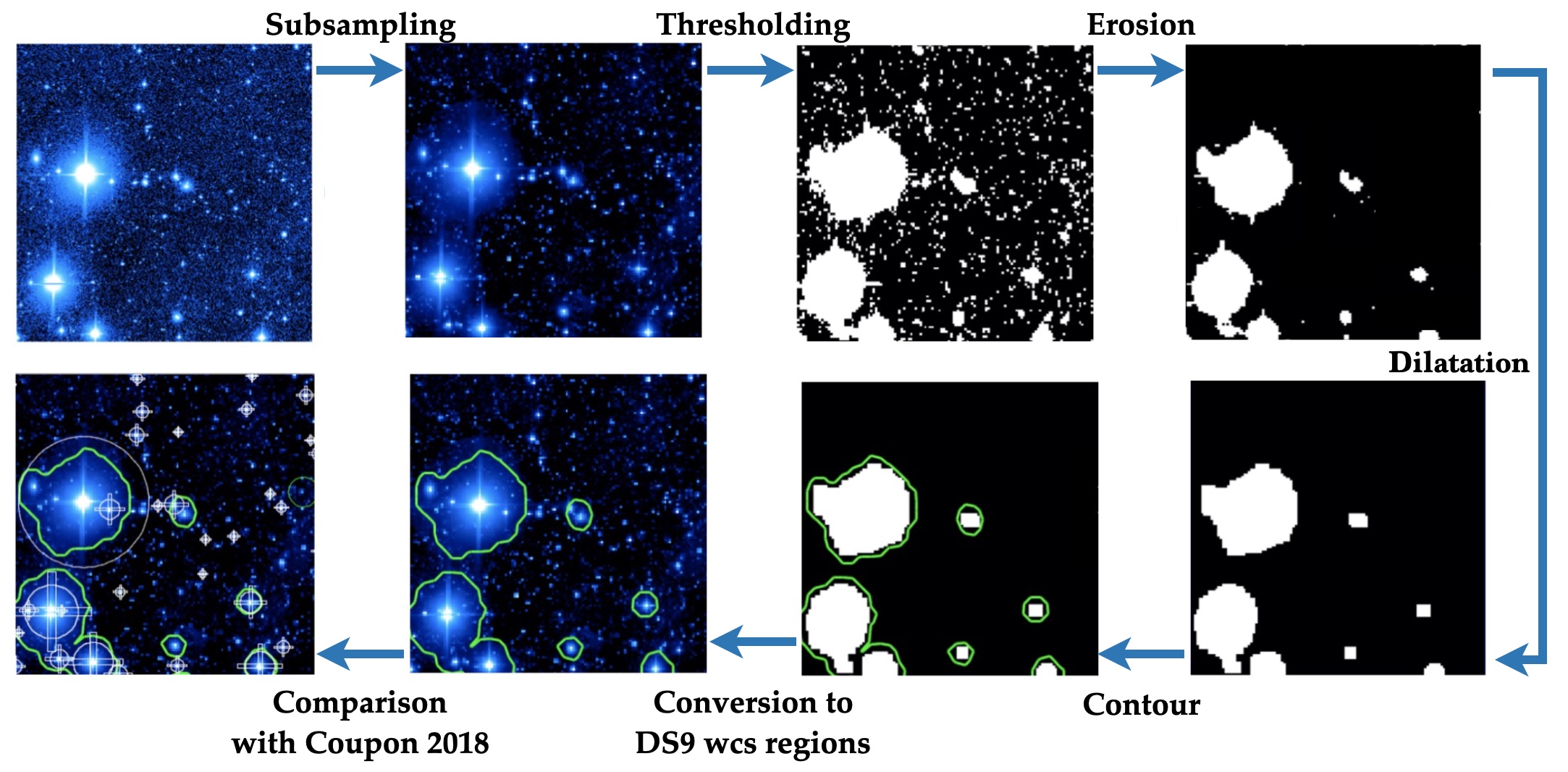}
     \caption{Flow chart and illustration of the process of mask creation. Images are smoothed and converted into binary masks with a threshold to identify bright extended objects. To avoid the masking of bright extragalactic objects, the masks are eroded (the pixels at the edges are set to 0), which removes small-scale objects. The masks are then dilated to recover the full size of the original masks for bright stars. Contours are created around masked areas (plotted in green) and we compare our new masks with the \cite{Coupon2018} ones (displayed as white circles and rectangles).} 
     \label{fig:Masks}
\end{figure*}

\citet{Aihara2019} identified some issues in the bright-star masks in the HSC PDR2, due to the change in the sky subtraction algorithm from HSC PDR1 \citep{Aihara2018b}. An outcome of this change is that the bright-star masks based on \textit{Gaia} and Tycho-2 catalogs presented in \citet{Coupon2018} are not conservative enough for PDR2. Moreover, stars in the NIR bands suffer more diffusion, implying that their halos are not completely covered by the masks. In addition to this problem, some bright stars are missing from the masks used by \texttt{hscPipe},\footnote{\url{https://hsc-release.mtk.nao.ac.jp/doc/index.php/bright-star-masks-2/}} requiring some improvements in the masking procedure.
To solve these two issues, we combine the original masks with new masks derived using  a simple  image-based approach that  follows the flowchart described in Fig.~\ref{fig:Masks}.
\texttt{Swarp} is used to smooth the images suffering the most scattering ($H$-band if available, HSC $y$-band otherwise) using the \texttt{LANCZOS3} method down to $\sim 5\arcsec$ pixel scale \citep{Bertin2002}. Then a threshold is applied to the pixel flux values to identify very bright objects and the images are converted onto binary masks (1 for masked and 0 otherwise) matching approximately the HSC ones. Finally, we apply to the binned masks a $5\times5$ erosion-dilation process. This process removes small masked areas and smooths the edges of the larger ones. %
Contours are converted into  \texttt{DS9} %
 polygon regions that can be handled by the \texttt{VENICE} code\footnote{\url{https://github.com/jcoupon/HSC_brightStarMasks}} \citep{Coupon2018} and applied to the source catalogs.

As shown in the comparison with the masks of \citet{Coupon2018} (bottom left plot of Fig.~\ref{fig:Masks}), our image-based approach does not mask all stars, but manages to mask the missing bright ($m_{ g}<16.5$) stars and increases the size of the mask that were undersized compared to the observations. The final bright-star masks are built by combining the \citet{Coupon2018} masks with those we generated.

\subsubsection{Satellite trails}

Despite the update of \texttt{HSC-Pipe} that identifies and clips transient artifacts before coaddition, PDR2 images (as well as CLAUDS images) still exhibit an important number of satellite trails ($\gg$10 deg$^{-2}$).
These trails could be easily handled via median-type stacking, but this would lead to a depth loss of 0.3 mag. An alternative is to use artifact rejection, but it misses a significant number of trails, which affects both source detection and photometry estimation. Indeed, \texttt{SExtractor} often fails at detecting sources close to bright satellite trails or generates spurious detections in those cases. This creates a bias in the number of counts per surface element. In addition, even when objects are detected around satellite trails, the photometry in the band with the satellite trail can be biased. This bias will propagate in the redshift estimation accuracy by generating catastrophic failures.

We decided to flag the regions possibly affected by these trails. We used  the convolutional neural network (CNN) \texttt{Maxi-Mask} \citep{Paillassa2020} to detect satellite trails in astronomical images. We used $0.6\%$ prior for satellite detections and thresholded the probability mapping to the minimum to get most of the trails. On 4200$\times$4200 images, $0.6\%$ corresponds to approximately two satellite trails of 10 pixels width crossing the image diagonally. \texttt{Maxi-Mask} can mismatch edge-on galaxies with track and generate false satellite detections that are significantly smaller than the satellite trails. We get rid of these false detections by removing any detection shorter than 1 arcminute.

We only applied the code to the detection images, as most bright satellite trails can be detected from them.
To avoid galaxies with photometry contaminated by the trails, we perform a 2\arcsec\ \ binary dilation on the satellite mask images and use \texttt{VENICE} to convert these binary images into the \textsc{ST\_TRAILS} flag in the released catalogs.

%


\section{Source detection and photometry}
\label{sec:photometry}
Source extraction and photometric analyses are two crucial steps to optimize the detection of faint sources in deep and crowded regions and to estimate the colors required for the photometric redshifts. As mentioned by \citet{Weaver2021}, photometric redshift estimates are more sensitive to the input flux catalogs than to the photo-$z$ codes themselves. Hereafter we perform these two steps with two different codes: the one implemented in \texttt{hscPipe} \citep{Bosch2018}, which has been adapted to handle images from other facilities (e.g., $U$ and NIR images), and the well tested and widely used \texttt{SExtractor} software \citep{Bertin1996}. 
 In the following part we describe the main steps followed by each software to generate the multiband catalogs and we compare the photometry obtained by the two catalogs as well as with the external catalog from COSMOS2020 \citep{Weaver2021}.

\subsection{\texttt{HSC pipeline}}
\label{sec:photometry-hsc}

The source detection process using the \texttt{hscPipe} is described in detail in \citet{Bosch2018}. It consists first of source detection in all individual bands. The configuration to run the detection on both U-band images is the same as that used in \citet{Aihara2019} on the HSC PDR2. However, for the NIR-data, a custom detection threshold was required to perform efficiently\footnote{Details on the detection configuration can be found here: \url{https://github.com/jcoupon/importExtData/blob/master/config/detectCoaddSources\_fixedThreshold.py}}, as the default configuration led to an important number of false detections of background fluctuations in these bands. Then, the lists of detected objects are merged and the photometry of all sources is extracted in all bands. Two measurements of the photometry are produced for the detected sources: one independent in all the bands, and one forced photometry across all the bands; we use the latter in the following sections. The full raw \texttt{hscPipe} data products, divided in tracts, patches (see Sect.~\ref{sec:data-preparation}), and bands, are available through the HSC data access portal (see Sect.~\ref{sec:data_access}). However, those data contain redundant sources 
due to the overlap between tracts and between patches that need to be cleaned to construct the final catalogs. In the E-COSMOS field, both U-bands were treated as two different sets of images, with detection and flux measurements made in both bands.

For each field, we 
include several photometric flux measurements: 2\arcsec\ and 3\arcsec\ aperture photometry, PSF photometry, Kron photometry \citep{Kron1980}, and cmodel photometry \citep{Abazajian2004}. Among these measurements both cmodel and PSF photometries take into account the PSF variation across patches and bands.
Quality flags from the output data are also included for each band to ensure the quality of the photometry. Sources identified by the \texttt{hscPipe} as duplicated are removed. 
The masks for bright stars used by the pipeline are the same as those used in \citet{Aihara2019} and thus suffer from the same issues as discussed in Sect.~\ref{sec:data-mask-brigh}. We thus apply the updated star masks presented in Sect.~\ref{sec:data-mask} as well as the satellite trail masks using the \texttt{VENICE} code \citep{Coupon2018}.

We check the consistency of our photometry with that obtained in \citet{Aihara2019} for the HSC PDR2 release.
Fig.~\ref{fig:hscpipe_vs_psdr2} in Appendix~\ref{sec:pdr2} shows the comparison for 2 arcsec aperture photometry. 
This comparison is done in the center of the E-COSMOS field where all the bands are available and where pipeline issues are most likely to happen, in particular in the processing of non-HSC data. The photometry of the \texttt{hscPipe} is in excellent agreement with the HSC PDR2 one. No bias is observed with magnitudes, and the scatter remains below $\sim$0.011~mag in all the HSC bands. 
This is also true for the cmodel magnitudes, where the scatter never exceeds 0.04~mag.

A compactness flag is computed in each band by comparing the PSF photometry to the cmodel one. A source that meets the following criterion: 
\begin{equation}
\label{eq:compactness}
{\rm mag}_{{\rm PSF},X}-{\rm mag}_{{\rm cmodel},X}<0.02
\end{equation}
is considered as compact in band $X$. If this criterion applies to all HSC bands, the source is considered as compact and flagged accordingly.
\subsection{\texttt{SExtractor} }
\label{sec:photometry-sextractor}
We also produce a version of the HSC-CLAUDS catalogs with the detection and photometry based on the \texttt{SExtractor} software \citep{Bertin1996}.

The source detection is performed on a detection image constructed by a nonlinear combination of the $U+grizy$ bands. When near-infrared observations are available, the detection image also includes the $K_s$ band.
 To compute the detection image we adopt the $\chi_{\rm mean}$ combination proposed by \citet{Drlica2018} as an alternative to the standard $\chi^2$ image \citep{Szalay1999}, because it minimizes the discontinuities between regions with a different number of input images. 
 
 The detection image, $\chi_{\rm mean}$, is defined as :
\begin{equation}
\label{eq:chi_mean}
\chi_{\rm mean} = \frac{\sqrt{\sum\limits_{i\le n} \frac{f_i^2}{\sigma_i^2}} - \mu}{\sqrt{n - \mu^2}},
\end{equation}
with 
\begin{equation*}
\mu = \sqrt{2}\frac{\Gamma((n+1)/2)}{\Gamma(n/2)}
\end{equation*}
and $f_i$ is the image in the band $i$, $\sigma_i^2$ its variance and n the number of input images used in the combination. The $\chi_{\rm mean}$ detection images are shown in Fig~\ref{fig:layout}. 

 Before constructing the $\chi_{\rm mean}$ images, we first checked the homogenization of the variance maps, as the procedures used to generate the images and their associated weight maps are different for the different datasets (CLAUDS, HSC, VIDEO or UltraVISTA). To ensure that the weight map properly scales as a variance map, we measured the standard deviation, $\sigma(\rm{Im})$, of 5000 randomly thrown 2\arcsec\ apertures in the background of the image, $\rm{Im}$, and compared with the median value, $\rm{median}(\rm{Var})$, obtained in its associated variance map,  $\rm{Var}$. For each patch and passband, we rescaled the variance map by the factor: $\sigma^2(\rm{Im})\ /\rm{median}(\rm{Var})$. 

For CLAUDS and HSC bands, the scaling  factors are around $\sim 2.5$--$3$, which means that without the rescaling we would underestimate the photometric errors by a factor $\sim1.5$. For the NIR data, the variance maps are rescaled by a factor $\sim 4$ in the deepest areas and up to $\sim 10$ in the shallowest areas, implying an underestimation of the photometric errors by a factor $\sim 2-3$. 

 The source detection is performed on the $\chi_{\rm mean}$ images convolved with a Gaussian with a 4$\times$4 pixel kernel. A source is detected if a minimum of 10 contiguous pixels is above a signal-to-noise ratio of 0.8$\sigma$ per pixel. The source deblending is controlled by the 
 \textsc{DEBLEND\_MINCONT} parameter, setting the fraction of the total intensity to be considered as a separated source, while the number of deblending thresholds is set by the \textsc{DEBLEND\_NTHRESH} parameter. We found that the values of 0.0003 and 64 for the two parameters provide efficient deblending.

 The detections and flux measurements are performed using the \texttt{SExtractor} dual mode. The $\chi_{\rm mean}$ image is used to detect and measure the shape parameters, while flux measurements are performed in individual images. 
 For each object, we provide two fixed aperture magnitudes (\textsc{MAG\_APER}) with 2\arcsec\ and 3\arcsec\ diameters and the Kron magnitude \citep[\textsc{MAG\_AUTO}][]{Kron1980}. 
 The Kron magnitude automatically adapts to the first-order moment of the light distribution, and its elliptical aperture is scaled to provide a pseudo-total magnitude \citep[]{Bertin1996}. However, fixed-aperture magnitudes are expected to be less noisy for faint sources and less impacted by blended sources, offering less noisy colors, a key input ingredient for the photometric redshift estimates \citep[]{Sawicki1997,Hildebrandt2012, Moutard2016}. As proposed by \citet[]{Moutard2016}, for each object, we compute a single offset (the same for all the bands) that allows us to convert from aperture to total magnitudes, that are required for physical parameters measurements:  
 ${m}_i = {m}_{{\rm APER},i} +\delta_m$, with the offset defined as:
\begin{equation}
\label{eq:offset}
\delta_m = \frac{1}{\sum\limits_i w_i} \times \sum\limits_i w_i.(m_{{\rm AUTO},i} - m_{{\rm APER},i})  ,
\end{equation}
where $i$ runs through the $Ugrizy$ and $J,K_s$ filters and $w_i$, the weight, defined as 
\begin{equation}
    \frac{1}{w_i} = \left(\frac{\sigma_{{\rm AUTO},i}}{f_{{\rm AUTO},i}} \right)^2  +
                    \left(\frac{\sigma_{{\rm APER},i}}{f_{{\rm APER},i}} \right )^2 ,
\end{equation}
where $f_\mathrm{AUTO}$ and $f_\mathrm{APER}$ are the fluxes associated to the Kron and aperture magnitudes respectively and $\sigma_\mathrm{AUTO}$ and $\sigma_\mathrm{APER}$ are their respective uncertainties. 
Being defined as a weighted average over all bands, the offset is more robust than if it were estimated for each band individually. However, some objects can have an unrealistic error in one band, which then accounts for more than $99\%$ of the total weight. To prevent this situation, we iteratively set to 0 the individual weights representing more than $95\%$ of the total weight.
The  offsets for the 2 and 3\arcsec\ aperture magnitudes are given in the final catalog. 
All the magnitudes provided in the catalog are corrected for galactic extinction based on the Schlegel extinction maps \citep{Schlegel1998}.

\subsection{Comparison and quality assessment}
\label{sec:photometrty-quality}
The HSC-CLAUDS photometry is assessed by comparing the measurements of our \texttt{hscPipe} and \texttt{SExtractor} catalogs. We also compare to a third and distinct method, delivered in the COSMOS2020 catalog \citep{Weaver2021}, based on \texttt{Farmer} software (Weaver et al. in prep.) which uses \texttt{The Tractor} \citep[]{Lang2016} to measure the photometry.
\texttt{Farmer} performs model photometry of individual sources with a simultaneous optimization of a group of nearby sources with additional constraints from the multi-channels. The COSMOS2020 catalog uses, among others, the same CLAUDS, HSC and UltraVISTA data as we did. The comparison is thus performed in the central E-COSMOS field, which is also the only region where all filters are available.  

\begin{figure}
    \centering
    \includegraphics[width=\linewidth]{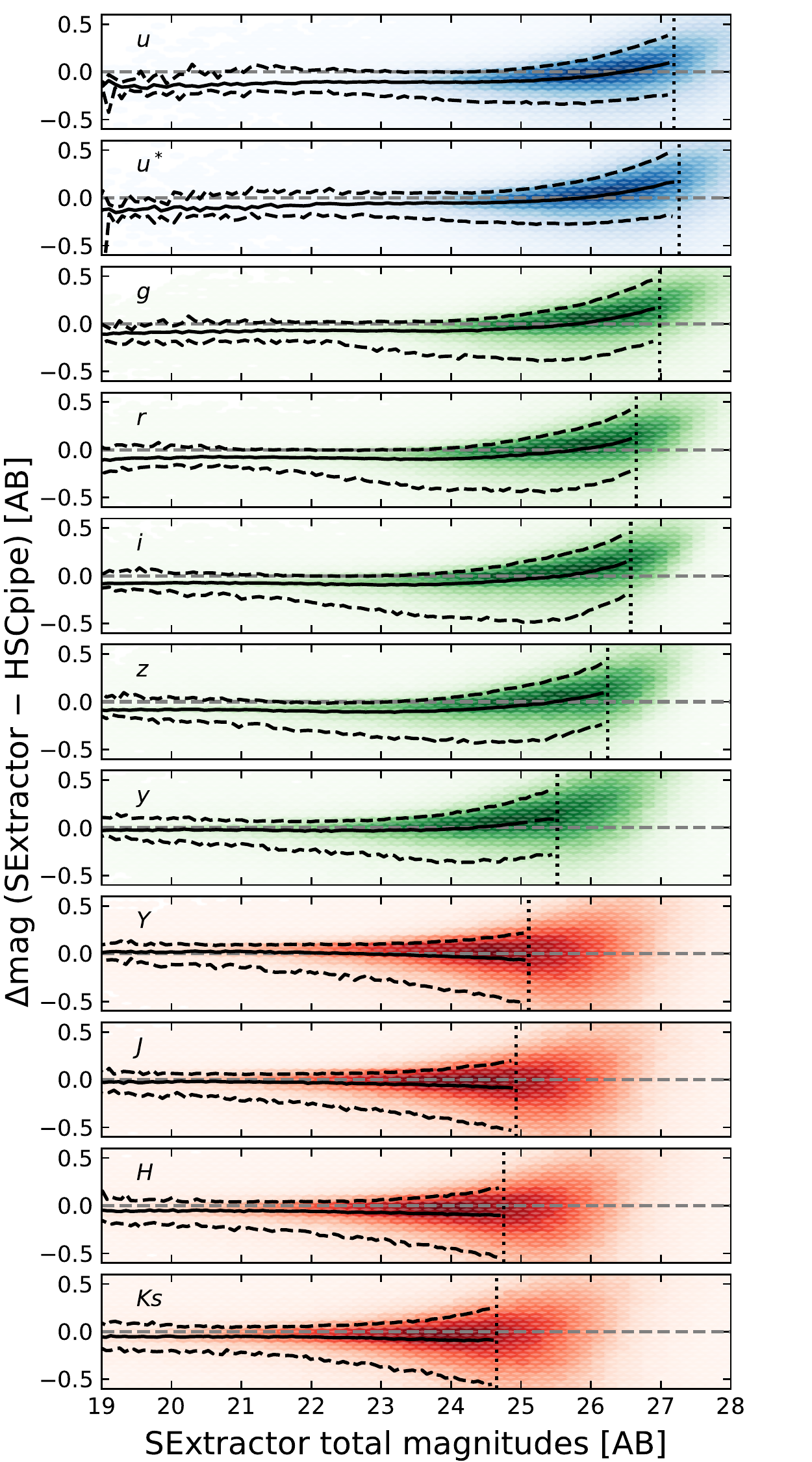}
    \caption{Comparison of \texttt{hscPipe} cmodel magnitudes with \texttt{SExtractor} total magnitudes. The color shades show the density of sources. The solid lines are the median of the distributions and the dashed lines are the $\pm34\%$ intervals around the medians. These values are computed down to the $5\sigma$ depth in each band in the E-COSMOS ultraDeep region (vertical dotted line)}
    \label{fig:comparisonSexHsc}
\end{figure}

\begin{figure}
    \centering
    \includegraphics[width=\linewidth]{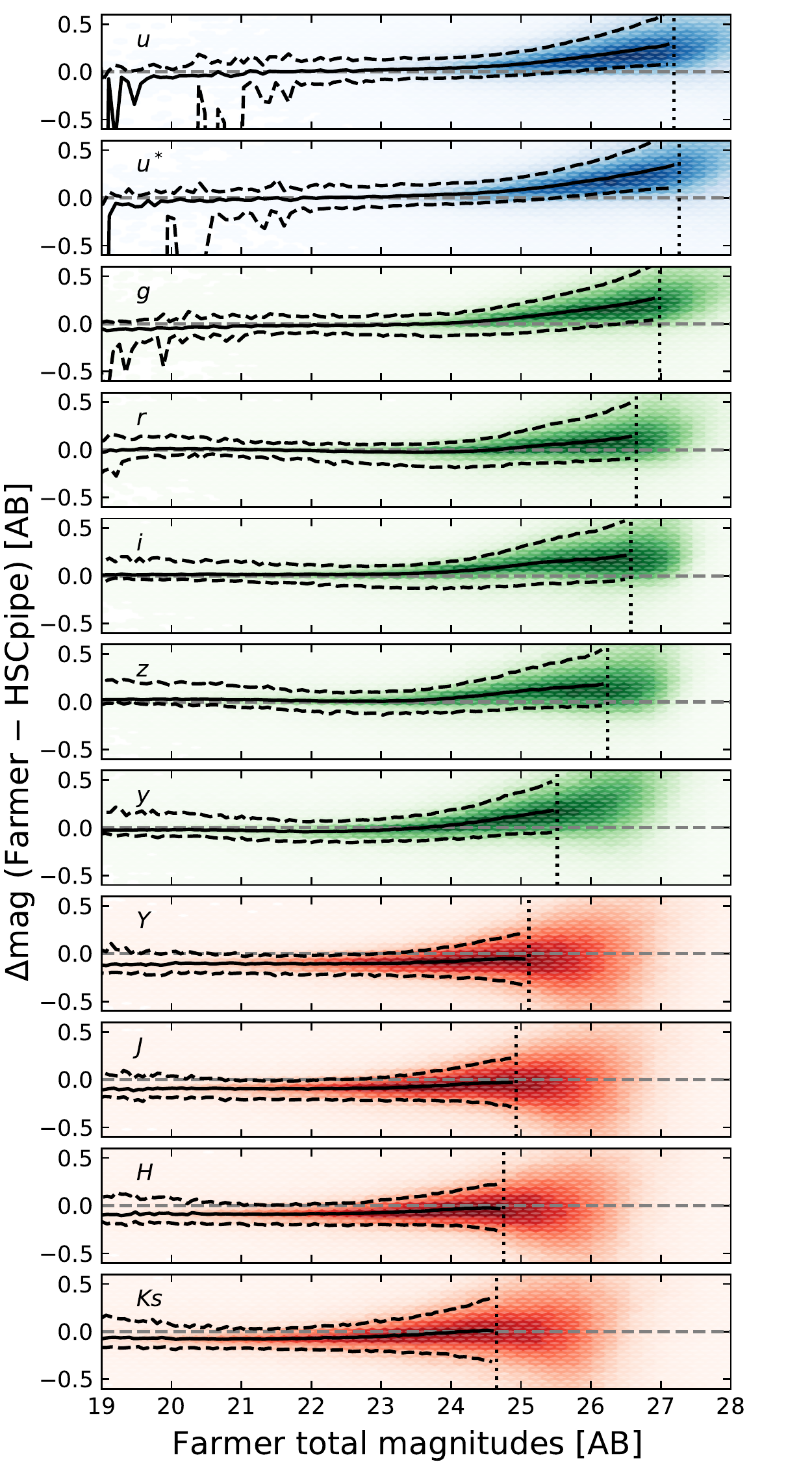}
    \caption{Same as Fig.~\ref{fig:comparisonSexHsc}, but comparing \texttt{hscPipe} cmodel magnitudes with COSMOS2020 \texttt{Farmer} total magnitudes. }
    \label{fig:comparisonFarmerHsc}
\end{figure}

\begin{figure}
    \centering
    \includegraphics[width=\linewidth]{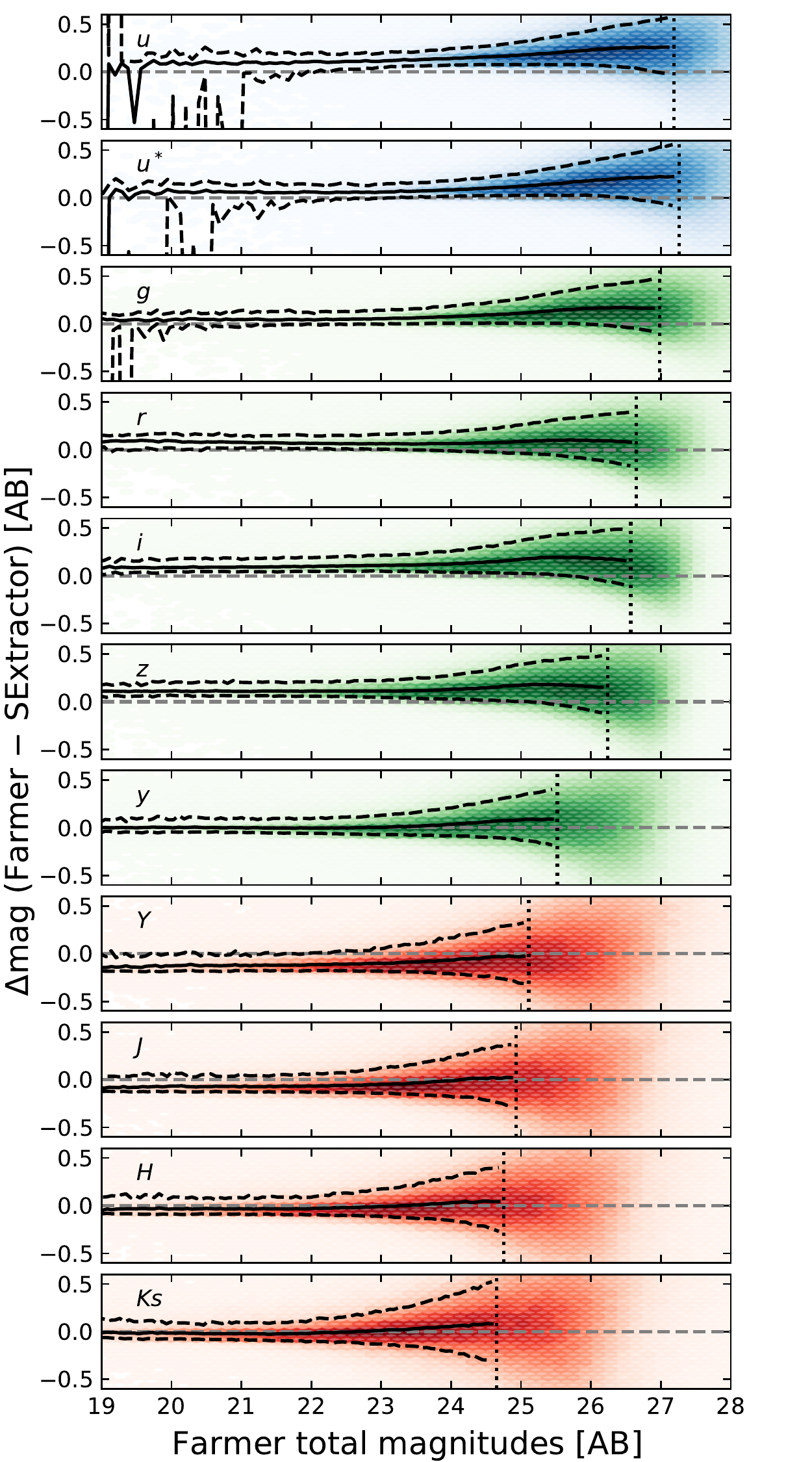}
    \caption{Same as Fig.~\ref{fig:comparisonSexHsc}, but comparing \texttt{SExtractor} with COSMOS2020 \texttt{Farmer} total magnitudes.}
    \label{fig:comparisonFarmerSex}
\end{figure}

In Figure~\ref{fig:comparisonSexHsc},  ~\ref{fig:comparisonFarmerHsc} and ~\ref{fig:comparisonFarmerSex} we compare the magnitudes.    
The cmodel from the \texttt{hscPipe} with the \texttt{SExtractor} aperture corrected magnitudes (Fig~\ref{fig:comparisonSexHsc}); the cmodel with \texttt{Farmer} magnitudes (Fig~\ref{fig:comparisonFarmerHsc}) and the  \texttt{SExtractor} corrected aperture with the \texttt{Farmer} magnitudes (Fig~\ref{fig:comparisonFarmerSex}).
The panels show for each filter the magnitude differences as a function of magnitude for the matched sources and the median of the distributions (solid lines) and their $\pm34\%$ intervals around the median (dashed lines) down to the $5\sigma$ depth measured in the ultraDeep region of E-COSMOS.
Several points are noteworthy. First, in the optical bands (bluer than $y$ band), we observe small offsets ($\le -0.1$ mag) in the bright end, with \texttt{SExtractor} magnitudes being systematically brighter than \texttt{hscPipe} and \texttt{Farmer} photometry, while almost no offset is observed between \texttt{hscPipe} and \texttt{Farmer}. The two latter magnitudes are based on modeled magnitudes and are not sensitive to seeing variations between bands, in contrast to the \texttt{SExtractor} aperture corrected magnitudes (see Eq.~\ref{eq:offset}), where no PSF homogenization has been applied, and the mean offset can be slightly over-estimated due to the worse seeing images. 

In the optical bands also, \texttt{hscPipe} magnitudes are getting gradually brighter than \texttt{SExtractor} and \texttt{Farmer} magnitudes for the faint objects (${m}\gtrsim 25.5$). The effect is much less pronounced between \texttt{SExtractor} and \texttt{Farmer} photometry. This faint-end trend could arise if the background correction in \texttt{hscPipe} is underestimated. This can be seen as well, although less pronounced, when comparing 2\arcsec\ aperture magnitudes between \texttt{SExtractor} and \texttt{hscPipe} optical magnitudes (Fig~\ref{fig:compSEx_HSC_2arcsec} in Appendix~\ref{sec:comparison2arcsec}). In contrast to the optical bands, in the near-infrared bands (redder than the $Y$ band),  the three different sets of magnitudes are consistent with only small systematic biases ($\le 0.05$) with no trend with magnitude. 

Finally, for all the bands, the distribution of the difference between the cmodel and \texttt{SExtractor} magnitudes shows a skewness with a large negative tail for the brighter \texttt{SExtractor} magnitudes. As a modeled photometry, the \texttt{hscPipe} may better handle the contribution of neighbors than the \texttt{SExtractor} aperture-corrected magnitudes. Consistently with this interpretation, the distributions for the two modeled magnitudes, \texttt{Farmer} and \texttt{hscPipe} (Fig~\ref{fig:comparisonFarmerHsc}), are almost symmetrical around the median, while the distributions between \texttt{Farmer} and \texttt{SExtractor} also show a slight asymmetry in the same direction.

\begin{figure*}
    \centering
    \includegraphics[width=\linewidth]{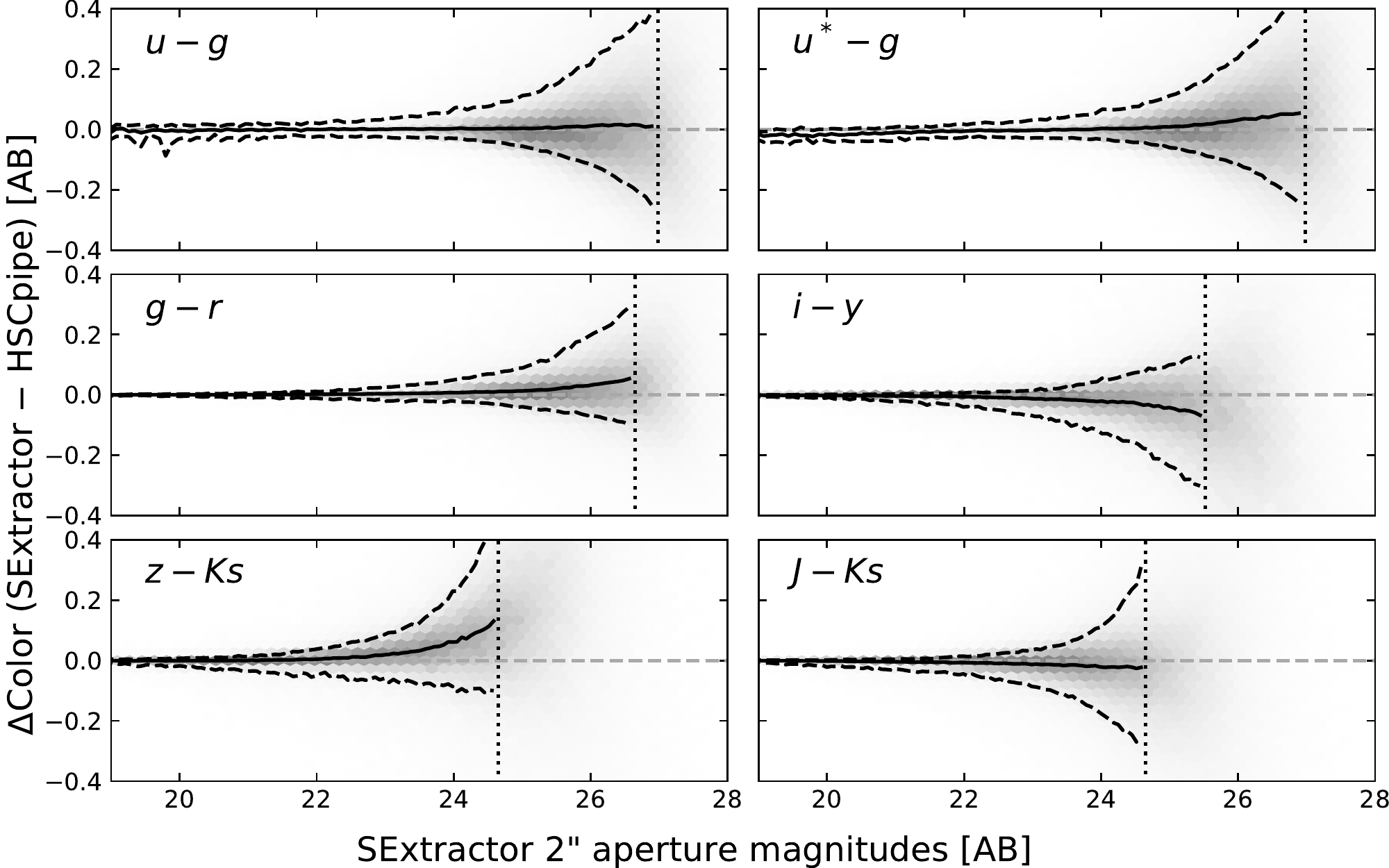}
    \caption{Comparison of \texttt{hscPipe} and \texttt{SExtractor} colors computed from 2\arcsec\ aperture photometry. In all the plots, the x-axis shows the \texttt{SExtractor} 2\arcsec\ aperture magnitudes of the second term in the colors. The solid line is the median of the distribution, and the dashed lines are the $\pm34\%$ confidence interval around the median. The vertical dotted lines mark the $5\sigma$ depth in the reference band (x-axis one).  }
    \label{fig:comparionColors}
\end{figure*}

%

Since the photo-$z$ computation is made using 2\arcsec\ aperture photometry with both catalogs (see Sections~\ref{sec:photoz-phosphoros} and~\ref{sec:photoz-lephare}), we make a comparison of the color apertures %
 in Fig.~\ref{fig:comparionColors}. We show the color differences between \texttt{SExtractor} and \texttt{hscPipe} photometry for six colors, namely 
$u-g$, $u^*-g$, $g-r$, $i-y$, $z-K_s$, and $J-K_s$. 
We note the good agreement between the colors provided by both catalogs up to magnitude $\sim 24$ in all the tested colors with no bias and a scatter $\lesssim0.1$. For fainter objects, a bias appears in colors involving optical and NIR bands, accompanied by a strong increase of the scatter. This can be due to the difference in background treatment, but also to the differences in depths. For $i-y$ and $z-K_s$ colors, this difference is $\gtrsim 1$~mag. The large scatter observed would be due to strong noise in the NIR band magnitudes of the faint objects compared to the optical ones.


\begin{figure}
    \centering
    \includegraphics[width=\linewidth]{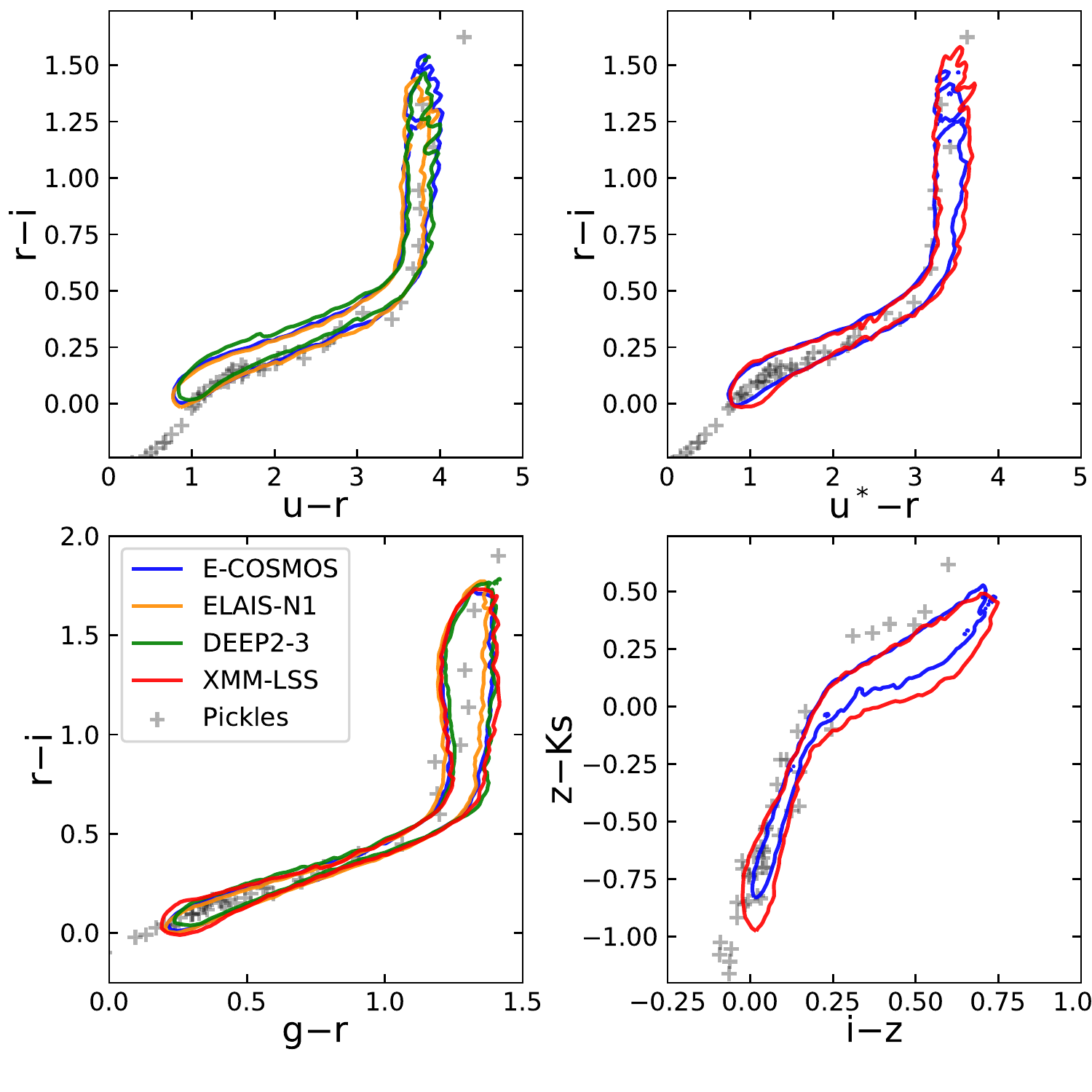}
    \caption{ Color-color density contours of stars in the different fields computed from the \texttt{hscPipe} PSF photometry. The contours enclose 68\% of the stellar loci. The crosses show the positions of main sequence stars from the Pickles library in these color-color diagrams. The offset between templates and observed loci indicates the need of proper extinction and zero-point corrections, which are applied in Sect.~\ref{sec:photoz} .}
    \label{fig:color_fields_hscpipe}
\end{figure}

To check the consistency of the photometric calibration between the four different regions of HSC-CLAUDS, we selected the stellar population and compare their stellar loci in color diagrams. We perform this test only with the \texttt{hscPipe} catalog, where a PSF magnitude is provided to get optimal stellar colors. 
Figure~\ref{fig:color_fields_hscpipe} displays the contours encompassing 68\% of the stars in four color-color diagrams for all the fields (color coded) available in a given color combination. 
The stellar loci for the different fields show an excellent overlap between each other in the optical bands. Combined with NIR color (bottom right panel, $z-K_s$ vs $i-z$), the stellar locus in the XMM-LSS field shows a larger scatter and a shift of $\sim 0.05$) compared to the E-COSMOS field.
The figure shows that the observed loci are not properly aligned with the main-sequence star colors computed from the Pickles template library \citep{Pickles1998}. This indicates a need for zero-point calibration to correct colors. From the diagrams, it appears that the $U$ bands and the NIR bands require the largest corrections. Although, for the bluest bands, the shift can also be due to Milky-Way extinction which is not accounted for because of the difficulty of computing it for objects (stars) inside the Galaxy.
We conclude that the photometry in the optical bands is consistent across the four fields but requires further correction in the form of zero-points calibration. The NIR band photometry might present some differences between E-COSMOS and XMM-LSS fields, which can be taken into account in the analysis by adjusting the zero-points on a field-to-field basis.\\

%

\begin{figure*}
    \centering
    \includegraphics[width=\linewidth]{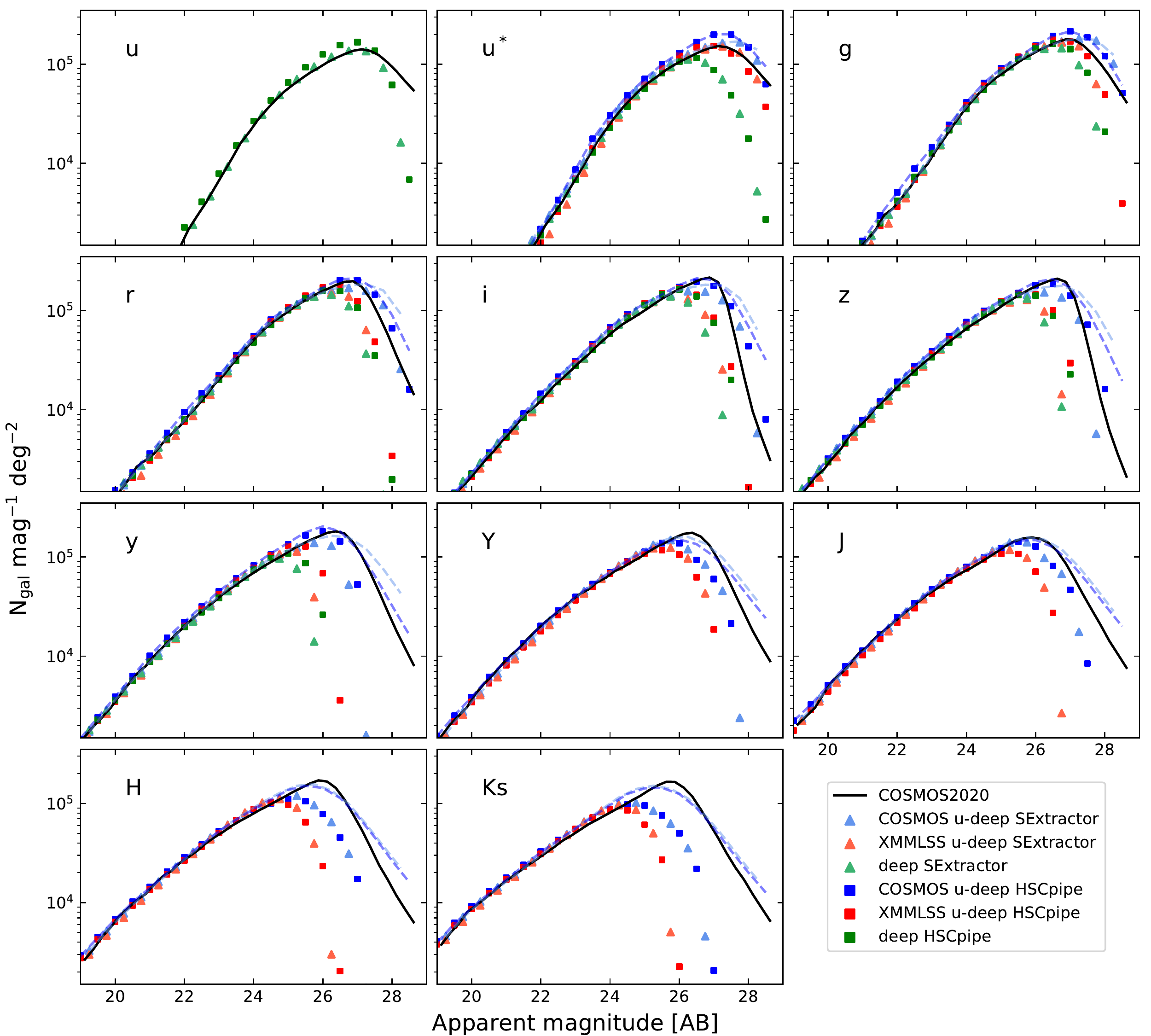}
    \caption{Number counts for \texttt{SExtractor} and \texttt{hscPipe} catalogs compared to COSMOS 2020 \texttt{Farmer} number counts. The number counts in the deep area combine the four fields. A cut at S/N$>3$ in all bands is made for \texttt{SExtractor} and \texttt{hscPipe} data displayed as points. The solid black line shows the number counts for COSMOS2020 \texttt{Farmer} photometry. The dashed lines show the \texttt{hscPipe} (dark blue) and \texttt{SExtractor} (light blue) number counts in the ultraDeep region of COSMOS without the cut in S/N.}
    \label{fig:numbercount}
\end{figure*}

%
Figure~\ref{fig:numbercount} shows the number counts in all bands for the \texttt{SExtractor} and \texttt{hscPipe} catalogs.  The number counts in the two ultraDeep regions (COSMOS-UD and XMM-LSS-UD) are presented separately, while those from the deep regions are the averages of the four fields.  
The comparison of the ultraDeep region of E-COSMOS and COSMOS2020 shows a good agreement, especially in the HSC bands, which is expected as the data used to build the catalogs are the same in the three cases. More differences arise in the NIR bands, first due to the difference in data, as COSMOS2020 uses the DR4 of UltraVISTA and our catalogs are built with the DR3, which is shallower. The number counts in Fig.~\ref{fig:numbercount} show the counts after applying a cut at S/N$>3$ in the concerned bands, which peak at brighter magnitudes than the COSMOS2020 counts. This could be explained by the rescaling of the errors, which affects strongly the NIR bands, leading to larger errors and hence pushing some objects below the S/N cut. The ultraDeep uncut counts match rather well with the COSMOS2020 ones, which is further evidence that the discrepancy is due to the cut in S/N.

For the bluest bands, there is a shift between the COSMOS2020 counts and ours. This shift could indicate some uncertainties in the estimation of the area that would produce a vertical shift, or systematic overestimation of the flux that would cause a horizontal shift, but it could also be explained by the differences in detection schemes used in the different pipelines. The \texttt{hscPipe} performs source detection in all the bands, while \texttt{SExtractor} uses a $\chi^2$ image constructed with bands from $U$ to $z$, with the addition of the $K_s$ band when available. The detection in COSMOS2020 uses only the reddest bands from $i$ to $K_s$. This difference makes our catalog more sensitive to blue objects, which explains why the discrepancy between the curves in Fig.~\ref{fig:numbercount} is stronger in the blue bands at faint magnitudes.

%


\section{Photometric redshifts}
\label{sec:photoz}
The photometric redshifts (photo-$z$s) are estimated with two template-fitting codes for the two photometric catalogs based on the 2\arcsec\ aperture flux in all bands. The \texttt{hscPipe} catalog is processed with the \texttt{Phosphoros} code, created to be part of the \textit{Euclid} photo-$z$ pipeline (Paltani et al., in prep.; \citealt{Desprez2020}), while the \texttt{SExtractor} catalog is processed with the well-tested code, \texttt{Le Phare} \citep{Arnouts2002,Ilbert2006}. 
As this is the first application of \texttt{Phosphoros} on deep imaging surveys destined to be released, we  also run \texttt{Phosphoros} on \texttt{SExtractor} photometry to distinguish the differences related to photometry issues from template-fitting code issues; this is detailed in Appendix~\ref{sec:PhosphorosSExtractor}.

For both codes, several configurations were considered and tested. We explored using different templates, prior information, photometry, and errors. The configuration presented in this section is the result of this exploration. We make the choice to use similar configurations, most of the differences being due to differences in the parameters both codes can adjust.

Both code use the templates library of \citet{Ilbert2013} for galaxies, consisting of a total of 33 spectral energy distributions (SED): seven elliptical galaxies, twelve spiral galaxies \citep{Polletta2007,Ilbert2009}, twelve starburst galaxies, and two elliptical SEDs generated with the \citet{Bruzual2003} stellar population synthesis models.
Intrinsic extinction is also configured in a similar manner in both codes.  Extinction was added as a free parameter for a reddening excess E(B-V) = 0, 0.05, 0.1, 0.15, 0.2, 0.25, 0.3, 0.4, 0.5 and by considering four attenuation laws: starburst-like \citet{Calzetti2000} and two dusty versions of the Calzetti law (with a bump at 2175$\AA$) for templates bluer than SB3, SMC-like \citet{Prevot1984} for templates redder than SB3 and no extinction for templates redder than Sb.

Emission lines are added to the templates by assuming an empirical relation between the UV light and H${\alpha}$ line flux \citep{Kennicutt1998} and ratios between the different hydrogen lines and oxygen lines  as measured from high redshift spectroscopic surveys \citep[]{Ilbert2009} or SDSS survey  (Paltani et al., in prep.).

As proposed by \citet[]{Ilbert2006}, to account for photometric variations in the four HSC-CLAUDS fields, the spectroscopic redshifts of bright sources ($19\le i \le21.5$) are used to apply a band-per-band zero-point correction in each field  separately. These corrections are given in Table~\ref{tab:zeropoints} for each field and combination of bands used. A small difference between the two configurations is that \texttt{Phosphoros} computed the zero points for both the $Ugrizy$ and $Ugrizy$+NIR configurations in the E-COSMOS and XMM-LSS fields.
Also, for both codes, the photo-$z$s computed in the central part of the E-COSMOS field use both $u$ and $u^*$ data separately when available. 

\begin{table*}[]
    \centering
    \caption{Compilation of the zero-point corrections computed by \texttt{Phosphoros} and \texttt{Le Phare}. The corrections are given in magnitudes, to add to the measured one. We chose that by convention the $i$-band correction would be null. For the \texttt{Phosphoros} computation, the corrections in the E-COSMOS and XMM-LSS fields were measured in both $Ugrizy$ and $Ugrizy$+NIR configurations.}
    \begin{tabular}{l cccccc|cccc}
    \hline
    \hline
    \rule{0pt}{1.2em}& \multicolumn{6}{c}{\texttt{Phosphoros}} & \multicolumn{4}{c}{\texttt{Le Phare}} \\
      & \multicolumn{2}{c}{E-COSMOS} & DEEP2-3 & ELAIS-N1 & \multicolumn{2}{c}{XMM-LSS}   &  E-COSMOS & DEEP2-3 & ELAIS-N1 & XMM-LSS \\
      \hline
      \rule{0pt}{1.2em}$u$   & \phantom{-}0.014 & \phantom{-}0.074 & -0.011 & -0.047 & -- & -- & -0.013 & \phantom{-}0.123 & \phantom{-}0.153 & --  \\ 
      
      $u^*$ & -0.111 & -0.048 & -- & -- & -0.150 & -0.030 & \phantom{-}0.110 & -- & -- & \phantom{-}0.131  \\
      
      $g$   & -0.011 & \phantom{-}0.046 & \phantom{-}0.017 & \phantom{-}0.015 & \phantom{-}0.003 & \phantom{-}0.071 & \phantom{-}0.019 & \phantom{-}0.102 & \phantom{-}0.073 & \phantom{-}0.033\\
      
      $r$   & \phantom{-}0.000 & \phantom{-}0.019 & \phantom{-}0.007 & -0.044 & \phantom{-}0.067 & \phantom{-}0.095 & -0.011 & \phantom{-}0.004 & \phantom{-}0.052 & -0.009\\
      
      $i$   & \phantom{-}0.000 &  \phantom{-}0.000 & \phantom{-}0.000 &  \phantom{-}0.000 & \phantom{-}0.000 &  \phantom{-}0.000 & \phantom{-}0.000 &  \phantom{-}0.000 & \phantom{-}0.000 &  \phantom{-}0.000  \\
      
      $z$   & \phantom{-}0.056 & \phantom{-}0.046 & \phantom{-}0.004 & -0.004 & \phantom{-}0.026 & \phantom{-}0.015 & -0.038 & \phantom{-}0.016 & \phantom{-}0.010 & -0.035 \\
      
      $y$   & \phantom{-}0.013 & -0.012 & -0.012 & -0.052 & \phantom{-}0.026 & \phantom{-}0.000 & \phantom{-}0.007 & \phantom{-}0.018 & \phantom{-}0.042 & \phantom{-}0.019\\
      
      $Y$   & -0.130 & -- & -- & -- & -0.096 & -- &  \phantom{-}0.156 & -- & -- & \phantom{-}0.161\\
      $J$   & -0.101 & -- & -- & -- & -0.110 & -- & \phantom{-}0.130 & -- & -- &  \phantom{-}0.140\\
      $H$   & -0.039 & -- & -- & -- & -0.026 & -- & \phantom{-}0.107 & -- & -- &  \phantom{-}0.131\\
      $Ks$   & \phantom{-}0.053 & -- & -- & -- &  \phantom{-}0.050 & -- & \phantom{-}0.020 & -- & -- &  \phantom{-}0.056\\
      \hline
    \end{tabular}
    \label{tab:zeropoints}
\end{table*}

\subsection{\texttt{Phosphoros}}
\label{sec:photoz-phosphoros}

\texttt{Phosphoros} %
is a template-fitting code %
that implements all the main functionalities of other template-fitting implementations. The focus on \texttt{Phosphoros} is the fully Bayesian treatment of the multi-dimensional likelihoods, with the possibility to apply complex priors and to marginalize over any parameters.
\texttt{Phosphoros} uses the configuration provided in Sect~\ref{sec:photoz}.

\texttt{Phosphoros} uses fluxes uncorrected for the Milky Way extinction, as the Galactic extinction is handled internally by \texttt{Phosphoros} by applying it directly on the fitted SED,  following the prescription of \citet{GalametzA2017}. 
Photometric errors in the different bands $x$, $\sigma_{f,x}$, are modified using the following equation:
\begin{equation}
    \sigma_{f,x}^{\rm new} = \sqrt{\alpha_x^2 \sigma_{f,x}^2 + \beta_X^2 f_{x}^2},
\end{equation}
where $f$ is the flux, $x$ the band and $\alpha$ is the uncertainty rescaling factor and is 2 for $Ugrizy$ bands and 4 for NIR bands, and $\beta$ corresponds to a systematic uncertainty (which could come from the photometry or from the models) and is set to 0.03 for all the bands. 

 We also consider several prior information.  When computing photo-$z$s using the NIR bands, a top-hat prior in absolute magnitude is applied, allowing only results with -24$<M_g<$-8, as well as a volume prior to take into account that the probability of a source to be in a slice of redshift increases with the volume of the slice.%
 When the number of available bands is restricted to the optical wavelengths, we apply a more informative prior based on luminosity functions from \citet{Zucca2006}. The rest-frame templates are sorted into four types according to their $B-I$ colors. For each type, we construct a prior assuming a Schechter function \citep{Schechter1976}, with the $\alpha$, the index of the power law at low luminosity,  and $\phi^*$, the source density, parameters given by \citet{Zucca2006} in the rest-frame $B$-band for galaxies between redshift $z=0.4$--$0.9$. However, the $M^*$ values are modified, setting them 3 magnitudes brighter, to avoid an overly strong suppression of luminous galaxies at high redshift. We introduce this modification because the strong evolution of $M^*$ makes the values determined by \citet{Zucca2006}, which are essentially obtained at low redshift, not applicable even at moderately high redshifts. Any small variation of this shift value ($\pm~0.5~{\rm mag}$) only has a small impact on the results. We prefer this prior to that on the magnitude-redshift distribution from \citet{Benitez2000} because its shape is physically better justified, which is especially important at high redshifts.

For each object, the probability distribution function of the redshift (PDZ) is obtained by marginalizing the posterior on the redshift dimension. The PDZs are defined with a fixed redshift grid ($z=0$--$6$) with a step $\delta z=0.01$. The photo-$z$s point estimates are defined as the median values of the PDZs. 
To deliver a consistent dataset across the entire HSC-CLAUDS survey, we provide a complete set of photo-$z$s with their PDZs based on the $Ugrizy$ photometry alone, and an additional set based on $UgrizyYJHK_s$ in the regions where near-infrared bands are available.

We also provide a star/galaxy separation by fitting the sources with stellar templates from the \citet{Pickles1998} library. In the catalog, all object with $\chi^2_{\rm star}<\chi^2_{\rm gal}$ are flagged as potential stars. Objects that are both potential stars and compact according to Eq.~\ref{eq:compactness} are flagged as stars.

\subsection{\texttt{Le Phare}}
\label{sec:photoz-lephare}

\texttt{Le Phare} \citep{Arnouts2002,Ilbert2006} is a well known and standard template-fitting method. %
With \texttt{Le Phare}, we adopted the same configuration as with \texttt{Phosphoros}, in terms of the SED library, attenuation laws and reddening excess, emission lines and zero-point calibrations (see above, section \ref{sec:photoz}). 

The SED fitting is run on \texttt{SExtractor} 2\arcsec\ aperture photometry, corrected for the Galactic extinction. 
Photometric errors are amplified by a factor 1.5, on top of which systematic magnitude errors of 0.02mag and 0.05mag are added in quadrature for optical and NIR bands respectively.
Depending on whether NIR photometry is used, a different prior set is adopted. In the case of $Ugrizy$+NIR,  a  $g$-band absolute magnitude is applied to reject solutions with $M_g<-24$. When only the optical bands are used, a fine-tuned prior similar to the \citet{Benitez2000} prior is used \citep{Ilbert2006}.

The redshift output of \texttt{Le Phare} is the median of the PDZ as point estimate, or, when the PDZ is not defined, the redshift is associated to the lowest $\chi^2$ value in the grid (the redshift grid is the same as the \texttt{Phosphoros} one, see Sect.~\ref{sec:photoz-phosphoros}). The photo-$z$s errors are computed from the 68\% confidence interval of the PDZ.

In addition to the galaxy template library, we also run the SED fitting with stellar and quasar template libraries.  The stellar library includes normal spectral types at solar abundance, metal-weak and metal-rich F-K dwarfs and G-K giants  \citep{Pickles1998},  low mass stars with different effective temperatures, gravity, and types \citep{Chabrier2000} and white dwarfs \citep[]{Bohlin1995}. The quasar library is a compilation of observed quasar templates \citep[]{Polletta2007} and templates with different contribution of galaxy and AGN spectra \citep{Salvato2009,Salvato2011}. 
 A first star/QSO/galaxy classification is provided, where compact objects with best fit obtained for a star or QSO template are flagged as such (see Appendix E).

\subsection{Comparisons}
\label{sec:photoz-comparison}
 \begin{figure*}
    \centering
    \includegraphics[width=\linewidth]{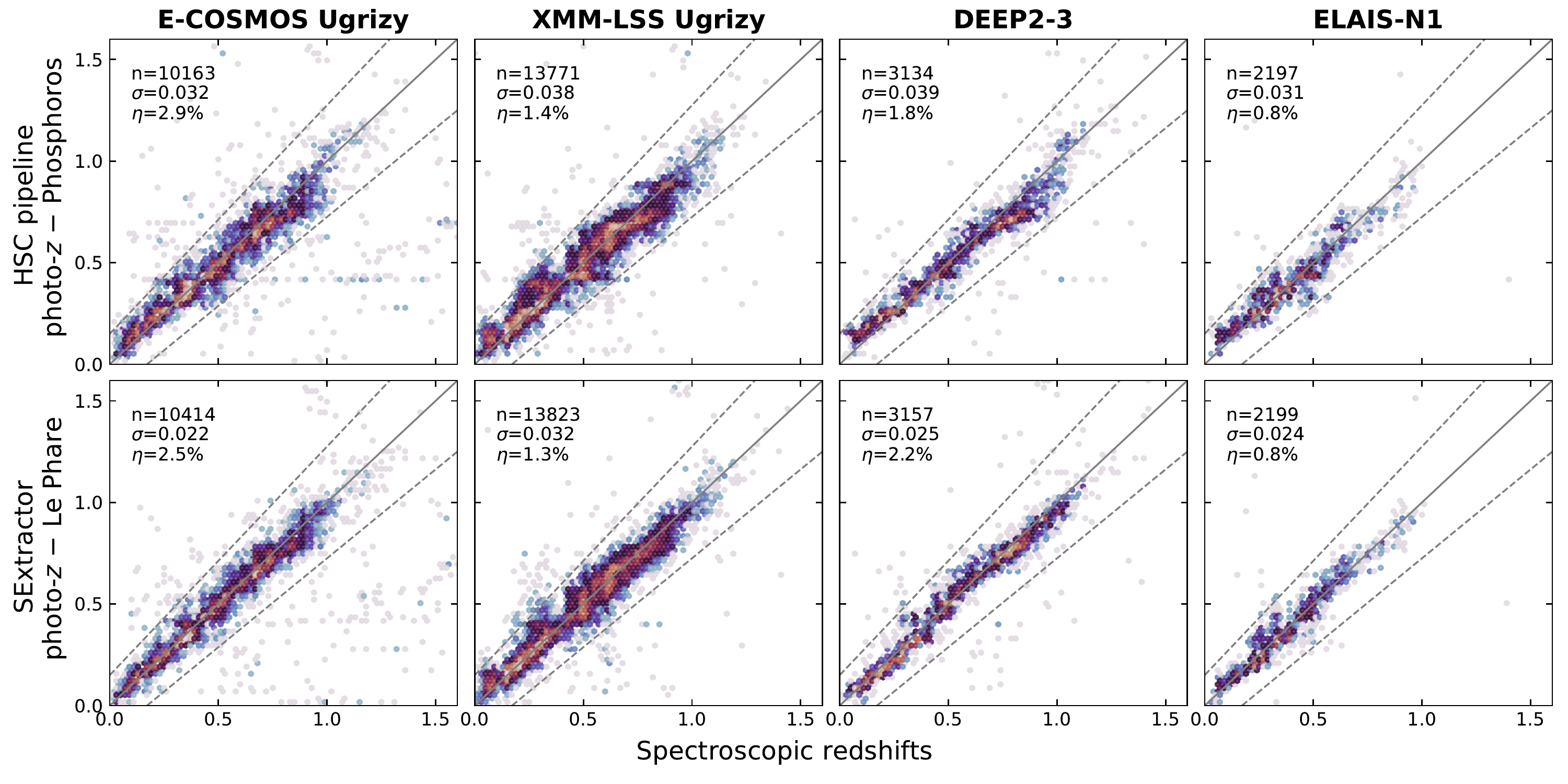} 
    \caption{Comparisons of the photo-$z$s based on the $Ugrizy$ photometry with spectroscopic redshifts down to $i\le 22.5$ in the four HSC-CLAUDS deep fields. Top panels show the photo-$z$s computed by \texttt{Phosphoros} code and the \texttt{hscPipe} catalogs, while the bottom panels show the   photo-$z$s computed with \texttt{Le Phare} and  \texttt{SExtractor} catalogs. 
    The solid gray lines show the 1:1 line and the dashed lines show the $z_{photo}=z_{\rm spec}\pm0.15(1+z_{\rm spec})$ corresponding to the limits for catastrophic failure.
    }
    \label{fig:photo-Z-i22.5}
\end{figure*}
In this section, we compare our photometric redshift estimates with spectroscopic redshifts (described in Sect.~\ref{sec:data-spectro}) and the latest 30-band photometric redshifts in the COSMOS field \citep[COSMOS2020, ][]{Weaver2021}.  
 As mentioned in \citet[][]{Weaver2021}, more than the differences in the photometric redshift codes, the main differences in the photo-$z$ vs spec-$z$ comparisons are due to the input photometric catalogs. This is especially true in our work where we use two SED fitting  codes with similar approaches but two quite different photometry codes. For this reason in the following we perform the comparisons using the photometric redshifts derived with the combinations \texttt{hscPipe}+\texttt{Phosphoros} (hereafter \texttt{HPH}) and  \texttt{SExtractor}+\texttt{Le Phare} (hereafter \texttt{SLP}). In Appendix~\ref{sec:PhosphorosSExtractor}, we show that the results from the two photo-$z$ codes are similar when run on  the same \texttt{SExtractor} catalog (see Fig.~\ref{fig:PhosphorosSExtractor}). 

\subsubsection{Metrics}
\label{sec:photoz-metrics}
The performance of the photometric redshifts is evaluated using  the following metrics: the residuals,  $\Delta z= (z_{\rm phot}-z_{\rm spec})/(1+z_{\rm spec})$, following the definition of \citet{Cohen2000};
the normalized MAD, $\sigma_{\rm MAD}=1.4826\times {\rm MAD}$, where MAD (Median Absolute Deviation) is the median of $|\Delta z - \textrm{Median}(\Delta z)|$ \citep{Hoaglin1983};
 and the fraction of outliers $\eta$ with $|\Delta z|\ge 0.15$.

 The photometric redshift point estimates are delivered with an uncertainty based on the 68\% confidence interval. To assess the relevance of the errors, for each galaxy $i$, we compute the absolute scaled residuals:
\begin{equation}
    D_{z,i} = \frac{|z_{{\rm phot},i}-z_{{\rm spec},i}|}{\delta^{68\%}_{i}},
    \label{eq:ASR}
\end{equation}
where $\delta^{68\%}_{i}$ is the 68\% confidence interval for the galaxy $i$. The cumulative distribution, $D_{z}$, should reach 0.68 for $D_z=1$.

We also assess the quality of the PDZs with the Probability Integral Transform (PIT) statistic \citep[PIT plot, ][]{Dawid1984,D'Isanto2018}. It is based on the shape of the distribution of the cumulative probabilities (CDF) at the true value: 
\begin{equation}
    C_{i}\equiv \mathrm{CDF}_{i}(z_{i}) = \int_{0}^{z_{i}} \mathrm{PDZ}_{i}(z)  \,{\rm d}z,
    \label{eq:CDF}
\end{equation}
for galaxy $i$ with spectroscopic redshift $z_i$ and probability distribution function ${\rm PDZ}_i$. If the spec-$z$’s can be randomly drawn from the PDZs then a flat PIT distribution is expected, suggesting that the PDZs are statistically correct, whereas convex or concave PIT distributions point to under- or over-dispersed PDZs, respectively \citep{Polsterer2016}. 

\subsubsection{Comparisons with spectroscopic redshifts}
\label{sec:photoz-comparison-specz}
Figure~\ref{fig:photo-Z-i22.5} presents density plots comparing the photo-$z$s from the \texttt{HPH} (top panels) and \texttt{SLP} (bottom panels) runs to spec-$z$'s for sources with $m_i<22.5$ in the four fields separately and based only on the $Ugrizy$ photometry. 
A zero-point calibration has been applied to each field separately.
The performance appears similar in the two configurations, with a precision varying between $\sigma=0.02$--$0.04$ and outlier fractions $\eta=2$--$3\%$ across the four fields.  However, we note a systematically higher scatter for the \texttt{HPH} results compared to the \texttt{SLP} ones by $\sim0.01$. \\
\begin{figure*}
    \centering
    \includegraphics[width=\linewidth]{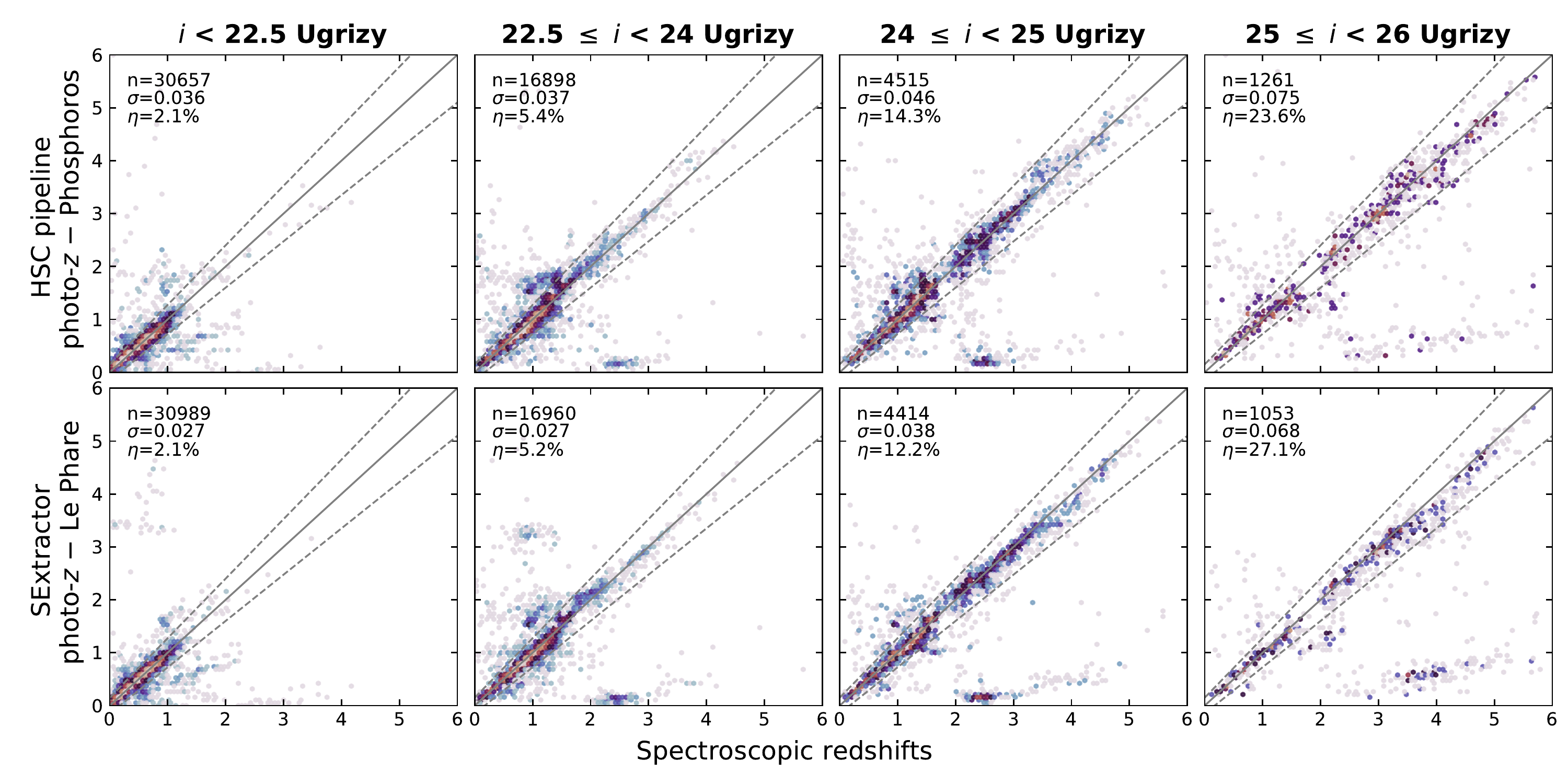}
    \caption{Same as Fig~\ref{fig:photo-Z-i22.5} but for all the HSC-CLAUDS fields combined and split by $i$-band magnitudes. The comparison is extended to faint magnitude bins, as indicated in each panel, with faint spectroscopic sources coming from the DEEP-23 ($m_i\le 24$), XMM-LSS, and E-COSMOS ($m_i\le26$) fields.
    }
    \label{fig:photo-z-bin-mag-i}
\end{figure*}
\begin{figure*}
    \centering
    \includegraphics[width=\linewidth]{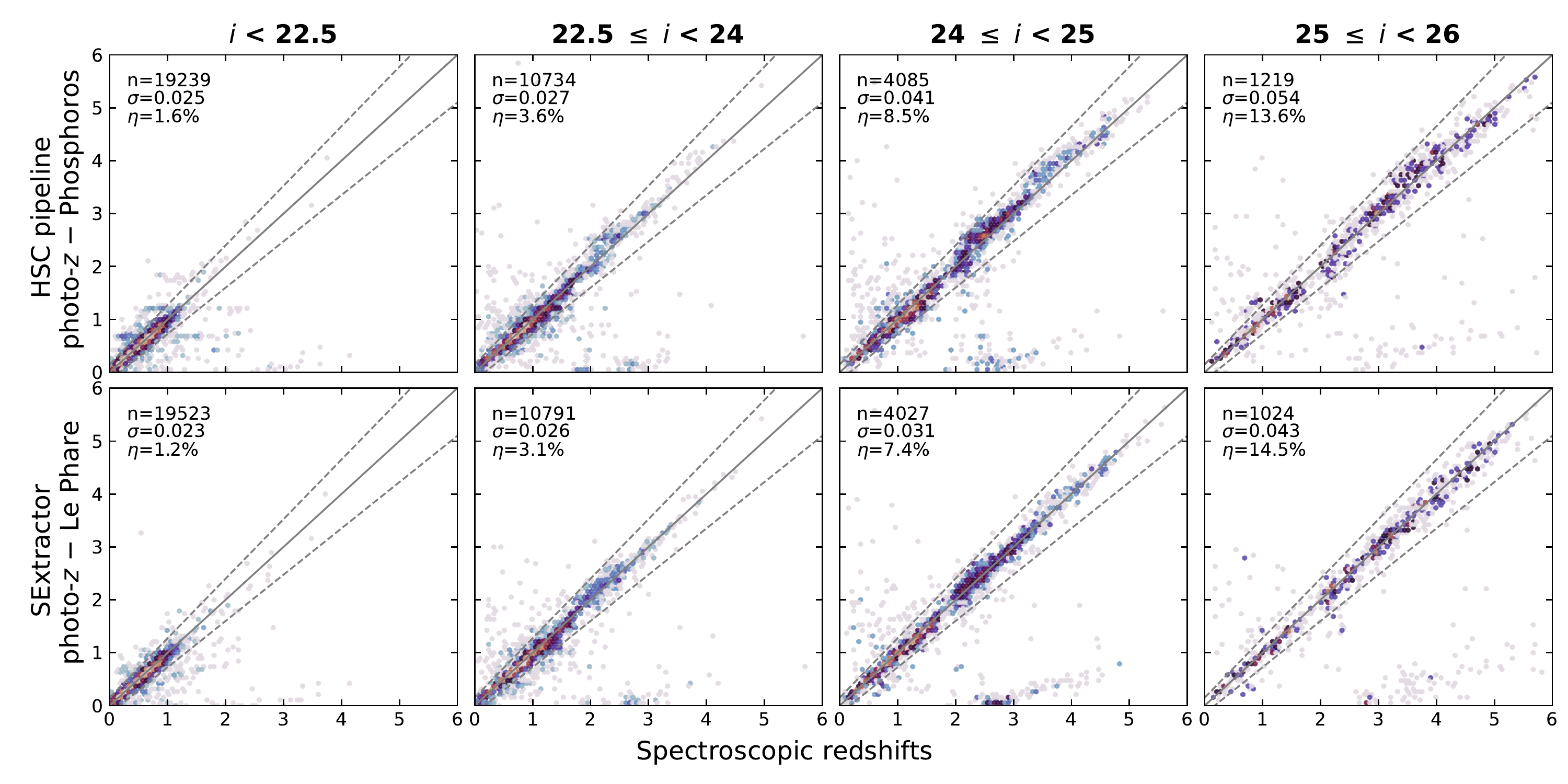}
    \caption{Same as Fig~\ref{fig:photo-z-bin-mag-i}, but for the XMM-LSS and COSMOS fields only and including the near-infrared photometry.
    }
    \label{fig:photo-z-bin-mag-i-nir}
\end{figure*}

In Fig.~\ref{fig:photo-z-bin-mag-i}, the four HSC-CLAUDS fields are combined together and the $Ugrizy$ photo-$z$s are compared with spectroscopic redshifts in faint magnitude bins. However, we note that the ELAIS-N1 field has almost no spectroscopic redshift with $m_i\ge 22.5$ and the DEEP-23 field only down to $m_i\le 24$. The vast majority of the faint spectroscopic sources comes from the XMM-LSS and E-COSMOS fields. \\
The precision of the photometric redshifts gradually deteriorates with magnitude, passing from $\sigma\sim 0.03$ at $m_i\le 22.5$ to $\sigma\sim 0.09$ at $m_i\le26$, with slightly worse results for the \texttt{HPH} implementation. Similarly, the fraction of catastrophic failures gradually increases from $\eta\sim 4\%$ to $\eta\sim 29\%$.
 
Although the fraction is similar between the two implementations, there are differences in the location of the outliers. Due to the Balmer and Lyman break confusion, two approximately symmetric clouds of outliers are generally produced, one for low spec-$z$ sources with a $z_{\rm phot}\geq 2$, and the other one for high spec-$z$ galaxies with a low photo-$z$s ($z_{\rm phot}\leq0.3$). The \texttt{SLP} implementation presents a visible cloud for sources with high spec-$z$'s and low photo-$z$s, while the \texttt{HPH} one leads to fewer sources in this cloud, but presents a higher population at low spec-$z$'s and high photo-$z$s than the \texttt{SLP} implementation. This difference is due to the application of a volume prior in the \texttt{Phosphoros} code since this volume prior favors the high redshift values for poorly constrained photometric sources.

A striking feature is present with spectroscopic sources around $z_{\rm spec}\sim 1$, to which higher photometric redshifts $z_{\rm phot}\sim1.5-2$ are assigned. The vast majority of those sources show a red $z-y$ color that can be interpreted either as caused by the 4000\AA\ break at $z\sim 1.5$, or by the contribution of emission lines (e.g., H\,$\beta$ and [O\,{\sc iii}] lines) at $z\sim 0.9$ (putting the 4000\AA\ break in the $r-i$ color). Without the near-infrared, this double solution cannot easily be broken by the $N(z)$ prior and occasionally leads to favor the wrong solution.           
 
In Fig.~\ref{fig:photo-z-bin-mag-i-nir} we show the comparisons of photometric redshifts measured by taking into account the NIR data. All the performance metrics are improved and we still observe slightly better results for the \texttt{SLP} implementation compared to the \texttt{HPH} one. 
For the faint galaxy population, the scatter and the catastrophic rate are improved by almost a factor of two. Including NIR data also alleviates the degeneracy observed around $z_{\rm spec}\sim 0.9$ with $z_{\rm phot}\sim 1.5-2$. This illustrates the benefit of the NIR observations to provide robust photometric redshifts over the entire redshift range.

\begin{figure}
    \centering
    \includegraphics[width=\linewidth]{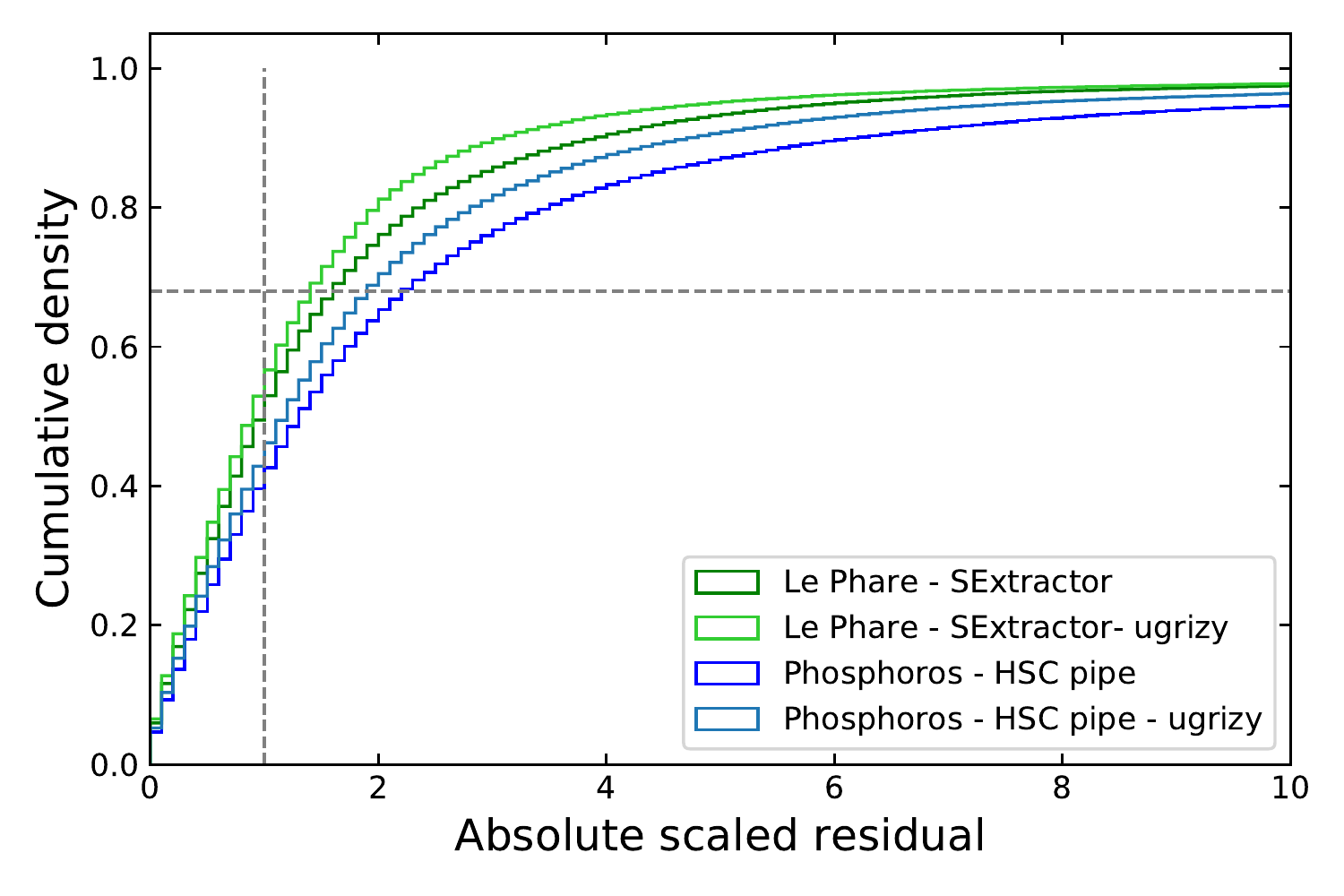}
    \caption{Absolute scaled residuals ($D_z$) cumulative distributions for the two implementations (HPH: blue lines and \texttt{SLP}: green lines) with the optical (light lines) and optical+NIR (dark lines) catalogs. The horizontal dashed gray lines show the 68\% level that all the distributions should cross at $D_z=1$ (vertical dashed line). 
    }
    \label{fig:ASR}
\end{figure}

The point estimate photometric redshifts are given with an uncertainty based on the 68\% confidence level interval derived from the individual redshift probability distribution function (PDZ).
To assess the relevance of the errors, for each source we compute the absolute scaled residual, $D_z$  (Eq.~(\ref{eq:ASR}), Sect.~\ref{sec:photoz-metrics}). Figure~\ref{fig:ASR} presents the cumulative distribution of $D_{z}$ for the two implementations
 (\texttt{HPH} and \texttt{SLP}) and the two imaging datasets (optical, optical+NIR).   
If the photo-$z$ uncertainties are correct we expect the cumulative curves to reach the 68\% level at $D(z)=1$. Indeed, all the cumulative distributions cross the 68\% level at higher $D(z)$, suggesting that the estimated photo-$z$ uncertainties are underestimated. In both configurations, the effect seems even worse when the near-infrared photometry is included, probably because of the additional constraints it provides on the PDZs.  Despite the fact that scaling factors have been applied to enlarge the photometric errors (see Sect.~\ref{sec:photoz-phosphoros} and~\ref{sec:photoz-lephare}), they seem insufficient to provide realistic photo-$z$ uncertainties. We note that \texttt{SLP} configuration seems to provide more reliable error estimates than \texttt{HPH}. As it is still the case when \texttt{Phosphoros} is run on \texttt{SExtractor} photometry, this difference is probably due to the code configurations, and in particular the rescaling of the errors, and not only to the photometry. We, therefore, caution that the current photometric-redshift uncertainties are on average underestimated. 

\begin{figure}
    \centering
    \includegraphics[width=\linewidth]{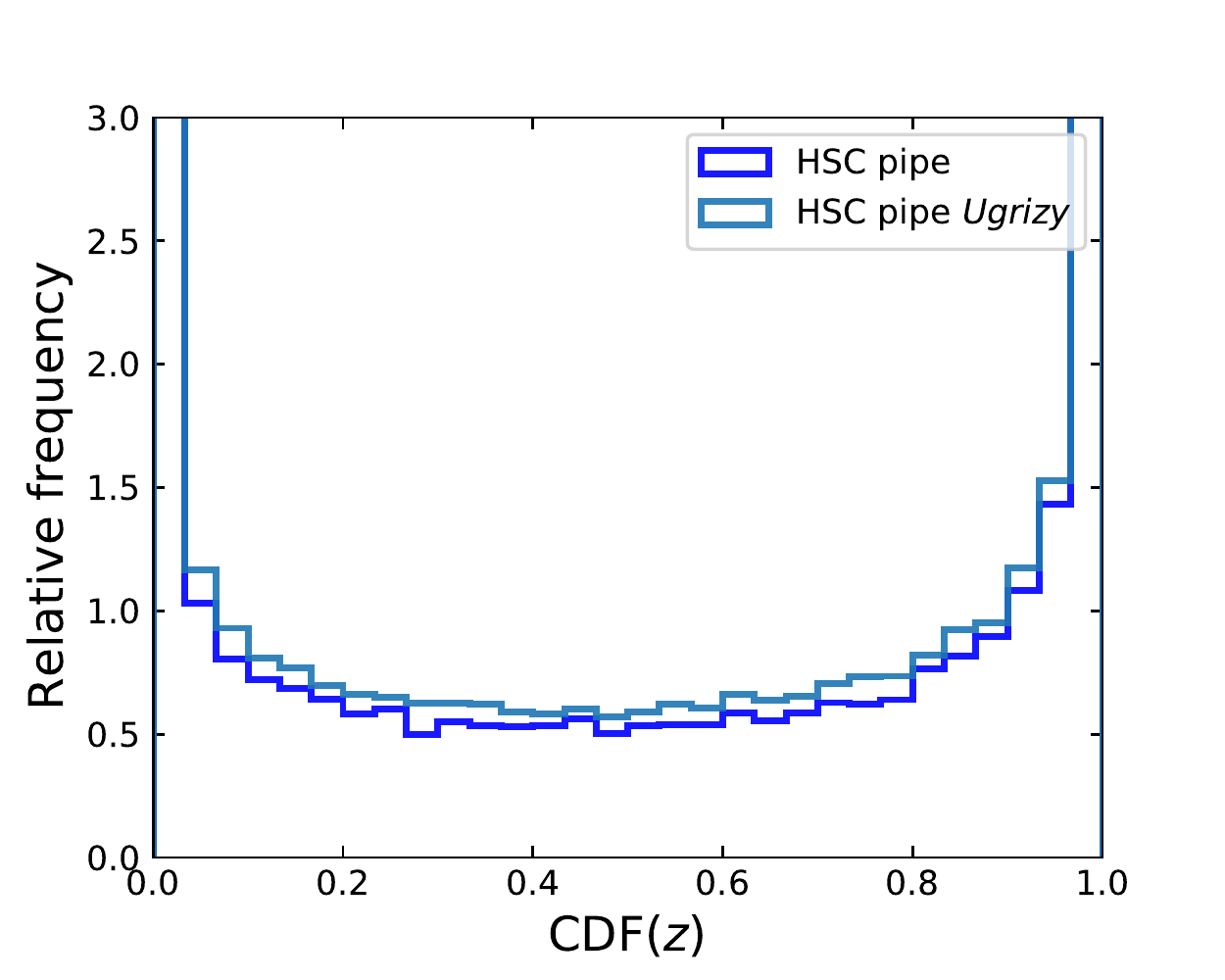}
    \caption{PIT plots for \texttt{Phosphoros} PDZs computed from the \texttt{hscPipe} photometry with (dark blue) and without (light blue) the infrared bands.}
    \label{fig:PIT}
\end{figure}

Alternatively, instead of the 68\% confidence interval, we can directly estimate the quality of the PDZs with a PIT plot
 (see Sect.~\ref{sec:photoz-metrics}).
 In Fig.~\ref{fig:PIT} we show the histogram of the \texttt{HPH} $C_{i}$ values.%
 The concave shape of the histograms shows that the PDZs are too narrow, leading to an underestimation of the errors. Such behavior has already been reported with several %
 template-fitting methods on similar data sets \citep{Tanaka2018,Desprez2020}. We also see that using the NIR data has the effect of constraining the PDZs even more. These observations are in agreement with the conclusions drawn from Fig.~\ref{fig:ASR}.

\subsubsection{Comparison with 3DHST and COSMOS2020 surveys}
At faint magnitudes, the spectroscopic samples are often selected with specific photometric criteria and are therefore not a representative sample of the overall photometric population \citep{Masters2015}. To address this issue, we compare our results to the 3DHST \citep[]{Skelton2014} and COSMOS2020 \citep[]{Weaver2021} surveys. The 3DHST is a slitless low-resolution grism survey that provides spectroscopic redshifts of near-infrared sources down to $m_H\le24$. However, 3DHST is still impacted by a selection bias due to the success rate of reliable redshift identifications, which is strongly dependent on the galaxy type, and in particular on the strength of the emission lines.
COSMOS2020 provides robust photometric redshifts over the entire color-space thanks to its 30 photometric bands, but these redshifts are less precise than the spectroscopic ones. These two samples are used to assess the quality of our photometric redshifts for the faint population. 
 
%
\begin{figure}
    \centering
    \includegraphics[width=\linewidth]{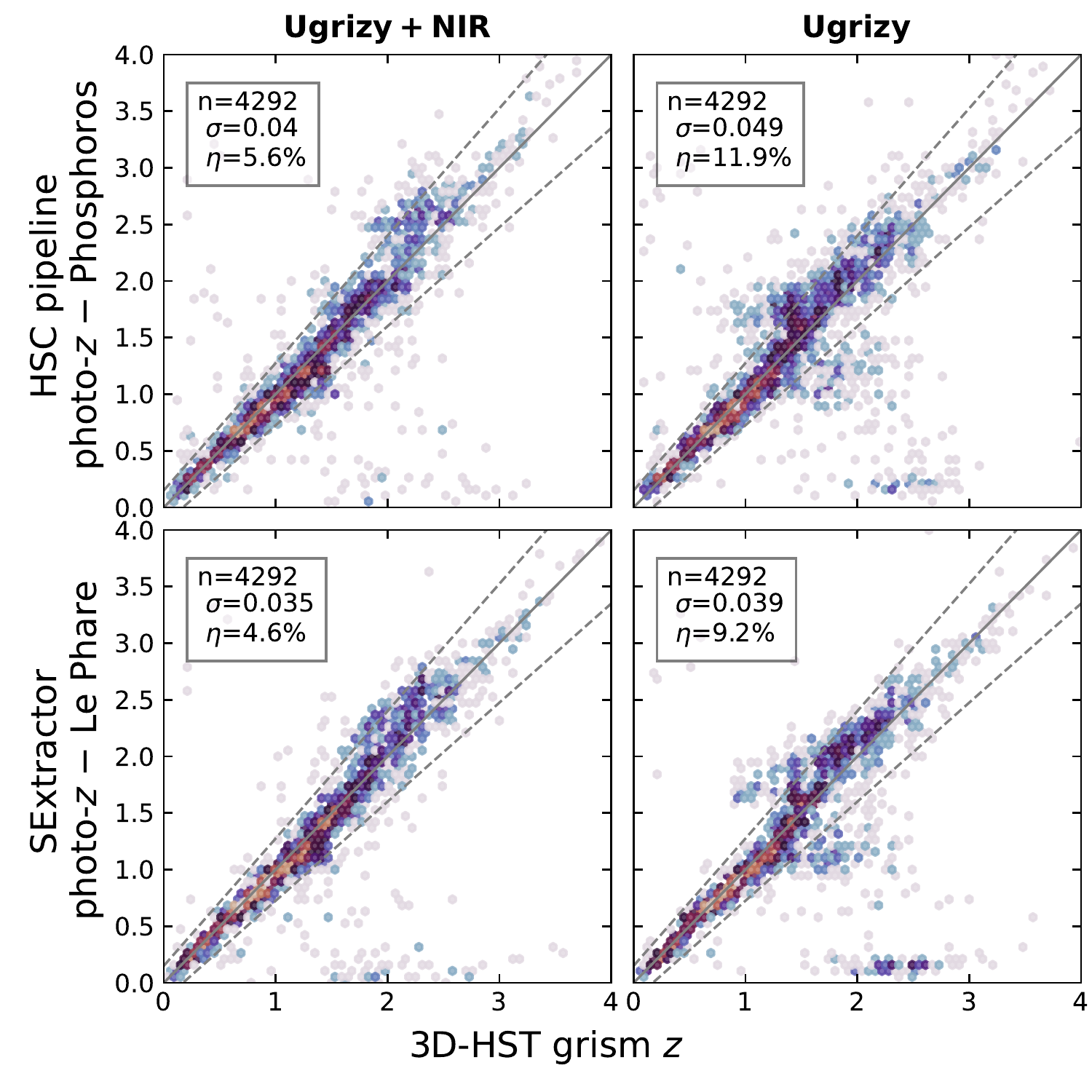}
    \caption{Photo-$z$ comparison with 3DHST redshifts for galaxies down to $m_H\sim 24$ for \texttt{HPH} (top panels) and \texttt{SLP} (bottom panels) catalogs, with the optical+NIR (left panels) and optical only (right panels) bands. The metrics are reported in each panel.}
    \label{fig:3dhst}
\end{figure}

  In Fig.~\ref{fig:3dhst}, we compare the photometric redshifts from the \texttt{HPH} and \texttt{SLP} %
  catalogs with the 3DHST redshifts, down to $m_H\sim$24. Two 3DHST fields overlap with the HSC-CLAUDS observations in the E-COSMOS and XMM-LSS fields and the sources are matched with a 1\arcsec\ tolerance.
 A small fraction of spectroscopic redshifts, especially when determined using slitless grisms, can be wrongly assigned; however, the impact should be negligible on the estimated scatter, as the median absolute deviation (MAD) is robust against outliers.
The results are consistent with the trends observed in Sect.~\ref{sec:photoz-comparison-specz} with optically selected spectroscopic sources. %
\texttt{HPH} photo-$z$s have slightly higher scatter and outlier fractions in the two adopted sets of filters. The catastrophic-failure fraction is reduced by a factor of two when including the NIR bands and the mismatch in the redshift domain $1\le z\le 2$ is noticeably reduced. On the code side, the implementation of priors used in \texttt{Phosphoros} for the optical configuration reduces the number of low photo-$z$ outliers compared to those from the \texttt{Le Phare} code.

\begin{figure*}
    \centering
    \includegraphics[width=\linewidth]{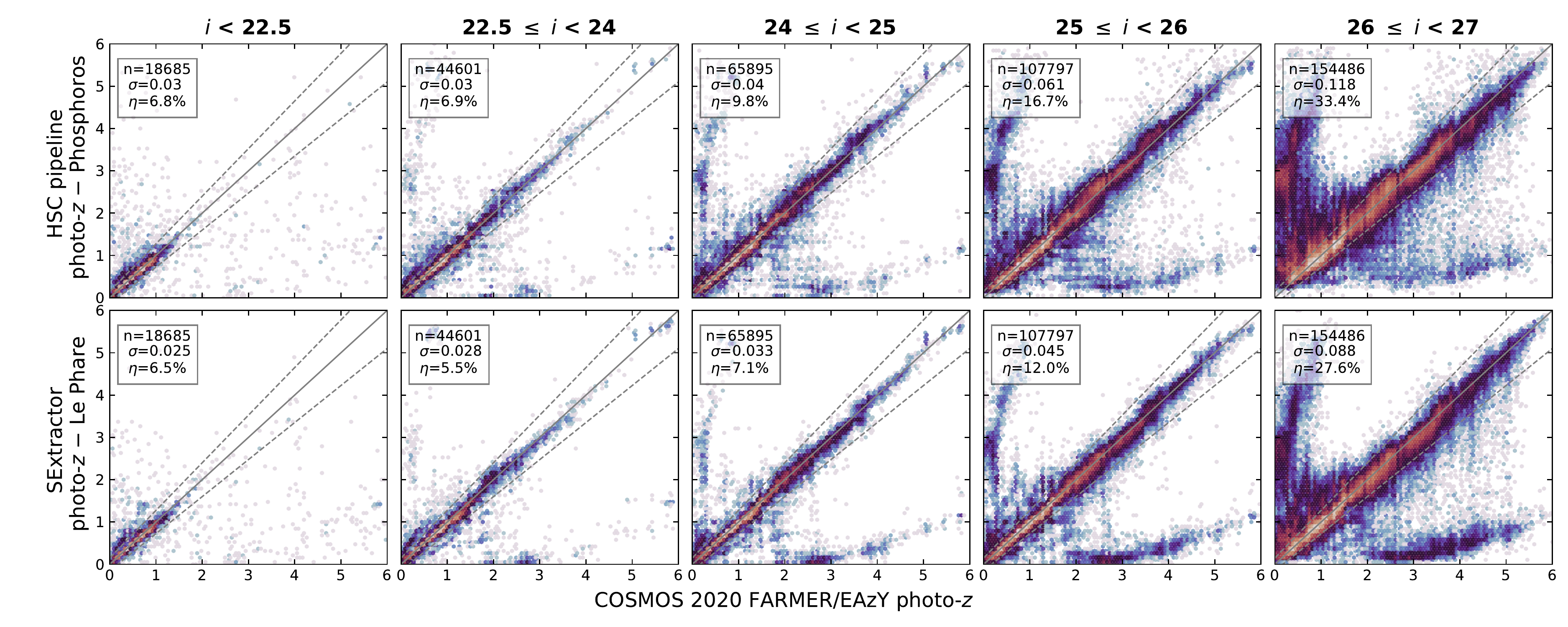}\\
    \includegraphics[width=\linewidth]{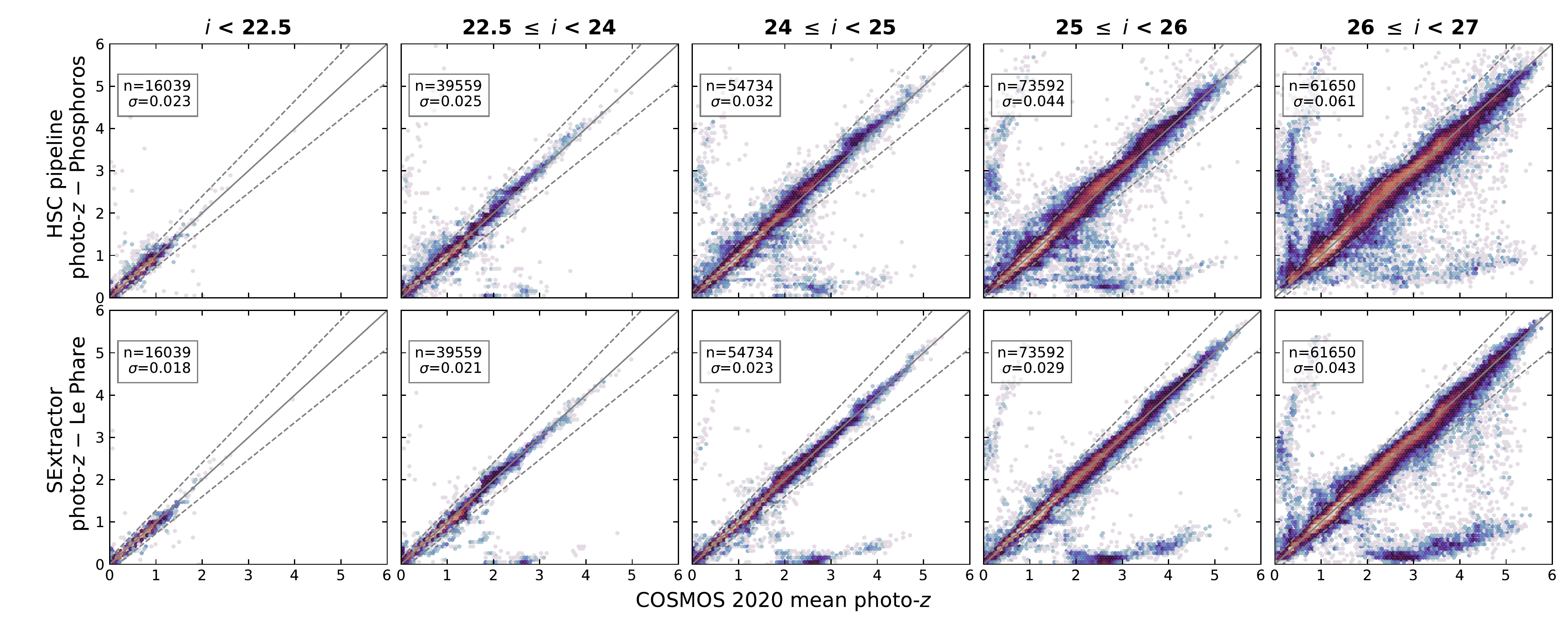}
    \caption{Comparison of \texttt{HPH} and \texttt{SLP} photo-$z$s to COSMOS2020 results. Top panels present the comparison with \texttt{Farmer}/\texttt{EAzY} results. The bottom ones show the comparison with the mean photo-$z$s computed from the four COSMOS2020 configurations clipped to only include photo-$z$s with standard deviation lower than $0.1[1+{\rm mean}(z)]$ between the four-point estimates. photo-$z$s have been computed using optical and NIR bands. The bins in $i$-band magnitude are based on the \texttt{FARMER} photometry.  }
    \label{fig:Photo-z-cosmos2020}
\end{figure*}

In Fig.~\ref{fig:Photo-z-cosmos2020}, we show the comparisons with the COSMOS2020 photometric redshifts. As described in \citet[]{Weaver2021}, four versions of photometric redshifts are available based on two photometric extraction softwares (\texttt{Farmer} and \texttt{SExtractor}) and two photometric redshift codes (\texttt{Le Phare} and \texttt{EAzY}; \citealt{Brammer2008}). In an attempt to be as independent as possible from source extraction and photometric redshift methods, in Fig.~\ref{fig:Photo-z-cosmos2020} (top 2 rows)
we first compare our measurements based on the optical+NIR bands with the COSMOS2020 \texttt{Farmer}/\texttt{EAzY} catalog.
However we point out that we use the same deep images (HSC, CLAUDS, and UltraVISTA) as COSMOS2020; these images provide the strongest constraints on the photo-$z$s at faint magnitudes.
But COSMOS2020 benefits from the ultradeep IRAC images and the narrow and intermediate Suprime-Cam bands \citep{Taniguchi2007,Taniguchi2015}. The former will reduce the number of catastrophic failures, while the latter will improve the scatter at bright magnitudes.    
In Fig~\ref{fig:Photo-z-cosmos2020}, the sources are restricted to cleaned flags (\textsc{FLAG$\_$COMBINED}=0) and sorted according to the \texttt{Farmer} $i$-band magnitude \citep{Weaver2021} to get the same number of sources in each magnitude bin.  
The comparisons show similar trends in the metrics to those in Fig.~\ref{fig:photo-z-bin-mag-i-nir} and~\ref{fig:3dhst}, with slightly better results for the \texttt{SLP} %
implementation. No systematic bias is observed up to redshift $z\sim 6$ and down to $m_i\sim26$ with a catastrophic-failure fraction below 15\%.  In the faintest magnitude bin, $26\le m_i\le 27$, the scatter almost doubles and the catastrophic-failure rate becomes significant, reaching 30\%; but those numbers are comparable to the level of consistency between the different photometric redshifts  in the COSMOS2020 catalogs \citep[][ see their Figure  14]{Weaver2021}.  
 The metrics are in good agreement with the values reported for the spectroscopic redshifts up to $m_i\sim 26$ (see Fig~\ref{fig:photo-z-bin-mag-i-nir}) and confirm that we do not expect major photometric-redshift issues for the whole galaxy population up to this depth. However two noticeable features are observed: first, a larger catastrophic-failure fraction at bright magnitudes with COSMOS2020 with respect to the comparison with spectroscopic redshifts; second, a mismatched photo-$z$ population at low redshift ($z_{\mathrm FARMER/EAzY}<0.5$) for which our two HSC-CLAUDS configurations estimate higher redshifts (above the 1:1 line). These two points are due to the fact that we are comparing our photo-$z$s with other photo-$z$s from COSMOS2020 that have their own inaccuracies.
To mitigate this effect, we make another comparison with the COSMOS2020 results by considering the four different versions of the photometric redshifts provided by \citet{Weaver2021}. For all the sources with four photo-$z$s, we compute the mean redshift ($\langle z \rangle$) and its standard deviation, $\sigma(z)$. We then select the COSMOS2020 self-consistent photo-$z$s by imposing a cut in the standard deviation $\sigma(z)<0.1(1+ \langle z \rangle)$. Strictly speaking, this does not constitute a guarantee to select secure redshifts, but rather it reduces issues related to either photometric extraction in problematic sources or sources with degenerate solutions, which may not be apprehended the same way by the two codes used in \citet[]{Weaver2021}. 

The comparisons with the mean COSMOS2020 photo-$z$s are shown in Fig.~\ref{fig:Photo-z-cosmos2020} (two bottom rows). With respect to the whole COSMOS2020 catalog, more than 85\% and 70\% of COSMOS2020 photo-$z$s are consistent with each other up to $m_i\sim25$ and 26 respectively. Our own photo-$z$s also appear consistent with the mean COSMOS2020 photo-$z$s with a scatter lower than $\sigma=0.03$ and a small fraction of catastrophic failures ($\eta\le 5\%$). 
 In the faintest magnitude bin, $26\le m_i\le27$, only 40\% of the COSMOS2020 redshifts are consistent with each other \citep{Weaver2021}. For this population, our photo-$z$s are still in good agreement with a scatter $\sigma\sim 0.05$ and a catastrophic-failure fraction not exceeding $\eta=10\%$.  
 As already noted, we observed with the COSMOS2020 photo-$z$s that the distribution of catastrophic failures for the \texttt{SLP} and \texttt{HPH} %
 catalogs are different. For the former, most of the catastrophic sources are high-z COSMOS2020 photo-$z$s which are assigned to low redshifts, while the latter mainly assign high redshifts to low-z COSMOS2020 photo-$z$s.  

\begin{table*}[]
    \centering
    \caption{Summary of the results in the different configurations tested in this work depending on $i$-band cuts. The \textit{weighted} column indicates if the results on the spectroscopic sample are weighted by the number of photometric sources (see Eq.~\ref{eq:weightsigma} and~\ref{eq:weighteta}). C~20 \texttt{F/E}  are the COSMOS2020 \texttt{FARMER/EAzY} photo-$z$s. }
    \begin{tabular}{l c c c  c c  c c c c c c c c}
    \hline
    \hline
    \rule{0pt}{1.2em}Sample& bands & weighted & \multicolumn{2}{c}{\texttt{$m_i<22.5$}} & \multicolumn{2}{c}{\texttt{$m_i<24$}} &\multicolumn{2}{c}{\texttt{$m_i<25$}} & \multicolumn{2}{c}{\texttt{$m_i<26$}} & \multicolumn{2}{c}{\texttt{$m_i<27$}}\\
    & & & $\sigma$ & $\eta$ & $\sigma$ & $\eta$ & $\sigma$ & $\eta$ & $\sigma$ & $\eta$ & $\sigma$ & $\eta$ \\
    \hline     
    \multicolumn{13}{c}{\rule{0pt}{1.2em}\texttt{hscPipe}/\texttt{Phosphoros}} \\
    \hline
    \rule{0pt}{1.2em}Spec-$z$& NIR &   & 0.025 & 1.6\% & 0.026 & 2.3\% & 0.027 & \phantom{0}3.1\% & 0.028& \phantom{0}3.4\% & \multicolumn{2}{c}{--} \\ 
    Spec-$z$& NIR & yes  & 0.025 & 1.6\% & 0.026 & 3.0\% & 0.034 & \phantom{0}6.0\% & 0.044 & \phantom{0}9.9\% & \multicolumn{2}{c}{--} \\ 
    Spec-$z$ &$Ugrizy$&  & 0.036 &2.1\% & 0.037 & 3.3\% & 0.037& \phantom{0}4.2\% & 0.038& \phantom{0}4.7\% & \multicolumn{2}{c}{--}  \\ 
    Spec-$z$ &$Ugrizy$&  yes & 0.036 &2.1\% & 0.037 & 4.3\% & 0.042 & \phantom{0}9.6\% & 0.057 & 16.1\% & \multicolumn{2}{c}{--}  \\      
     C~20 \texttt{F/E} & NIR  & & 0.030 & 6.8\% & 0.030 & 6.8\% & 0.035 & \phantom{0}8.3\% & 0.045 & 12.2\% & 0.063 & 20.5\%  \\ 
     C~20 \texttt{F/E} & $Ugrizy$  & & 0.036 & 8.0\% & 0.038 & 9.4\% & 0.044 & 12.6\% & 0.055 & 17.1\% & 0.079 & 25.8\%  \\
    \hline
    \multicolumn{13}{c}{\rule{0pt}{1.2em}\texttt{SExtractor}/\texttt{Le Phare}} \\
    \hline
    \rule{0pt}{1.2em}%
    Spec-$z$& NIR & &  0.023 & 1.2\% & 0.024 & 1.9\% & 0.025 & \phantom{0}2.5\% & 0.026 & \phantom{0}2.9\% & \multicolumn{2}{c}{--} \\ 
   Spec-$z$& NIR & yes &   0.023 & 1.2\% & 0.025 & 2.5\% & 0.028 & \phantom{0}5.1\% & 0.035& \phantom{0}9.5\% & \multicolumn{2}{c}{--} \\ 
    Spec-$z$ & $Ugrizy$&  & 0.027 & 2.1\% & 0.027 & 3.2\% & 0.028 & \phantom{0}4.0\% & 0.028& \phantom{0}4.4\% & \multicolumn{2}{c}{--} \\ 
    Spec-$z$ &$Ugrizy$ & yes  & 0.027 & 2.1\% & 0.027 & 4.2\% & 0.033 & \phantom{0}8.4\% & 0.049& 17.2\% & \multicolumn{2}{c}{--} \\ 
     C~20 \texttt{F/E} &  NIR &  &  0.025 & 6.5\% & 0.027 & 5.8\% & 0.030 & \phantom{0}6.5\% & 0.036 & \phantom{0}9.0\% & 0.049 & 16.3\%  \\ 
     C~20 \texttt{F/E} &  $Ugrizy$ &  &  0.026 & 6.9\% & 0.028 & 6.8\% & 0.034 & \phantom{0}8.5\% & 0.042 & 11.7\% & 0.060 & 20.7\%  \\ 
    \hline
    \end{tabular}
    
    \label{tab:metrics_summary}
\end{table*}

In Table~\ref{tab:metrics_summary} we report the scatter and the catastrophic-failure fractions for the different configurations for different $i$-band cuts tested against the spectroscopic sample and COSMOS2020. In the former case, the scatter and outlier fraction values are misleading, because the spec-$z$ sample is strongly biased toward brighter magnitudes compared to the whole photometric catalog. Therefore we provide scatter $\sigma$ and outlier fraction $\eta$ values weighted by the number of photometric objects in each magnitude bin, which provides a much more realistic view of the performance on the full photometric sample \citep{Desprez2020,Hartley2020}. There is no such problem when comparing with COSMOS2020 since it contains essentially all photometric sources. The weighted $\sigma$ and $\eta$ for the spec-$z$ sample are thus computed as follows:
\begin{equation}
    \label{eq:weightsigma}
    \sigma_{\rm weighted} = \frac{1}{\sum_{j} n_j}\sum_{j} \sigma_j n_{j},
\end{equation}
and:
\begin{equation}
    \label{eq:weighteta}
    \eta_{\rm weighted} = \frac{1}{\sum_{j} n_j}\sum_{j} \eta_j n_{j},
\end{equation}
where $j$ represents the magnitude bins ($m_i<22.5$, $22.5\leq m_i<24$, $24\leq m_i<25$, etc.), $n_j$ is the number of sources in the photometric catalog in the magnitude bin $j$, and $\sigma_j$ and $\eta_j$ are the metrics for bin $j$. The weighting scheme doubles the outlier fractions measured in the spectroscopic sample with a $m_i<25$ cut and triples it at $m_i<26$.  
We note that the scatter values with the weighted scheme are consistent with those of COSMOS2020 \texttt{FARMER}/\texttt{EAzY}. For the outlier fractions, we note some differences especially for the bright cuts, as by construction the results in the first magnitude bin do not change. It is with a cut at $m_i<25$ that we have the most consistent results. With this cut, we have a scatter $\sigma\lesssim0.04$, even when using only $Ugrizy$ data and after weighting the spectroscopic sample results or using the COSMOS2020 data as a reference. With the same cut, we obtain an outlier fraction $\eta<0.05$ for the spectroscopic sample, regardless of the photometry used. At $m_i<25$, the catastrophic-failure rate remains below 10\%, both when compared with the spec-$z$ sample and with COSMOS2020.

%

%
\subsubsection{Redshift distributions}
\label{sec:photoz-comparison-nz}
\begin{figure*}
    \centering
    \includegraphics[width=\linewidth]{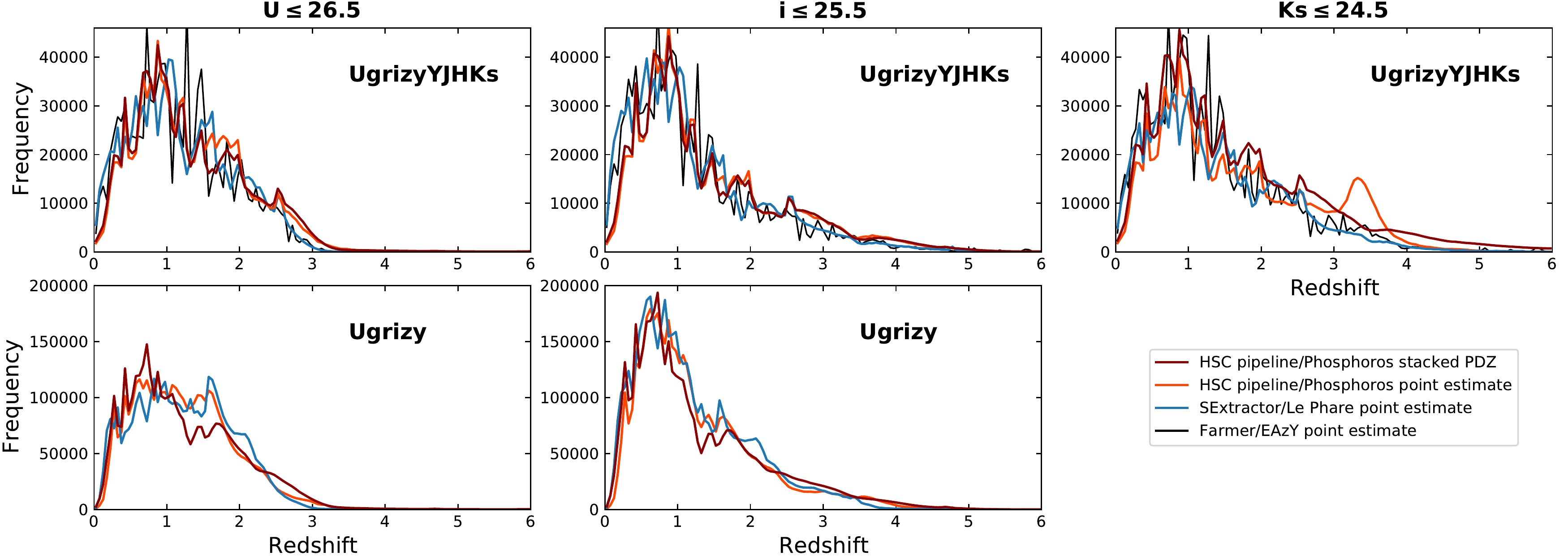}
    \caption{Galaxy redshift distributions ($N(z)$) in the CLAUDS-HSC regions for three selections, as indicated in the top of each panel. The three panels in the top row show the results based on estimates with all the bands and are thus limited to E-COSMOS and XMM-LSS fields. The two panels in the bottom row show those computed with $Ugrizy$ photometry for the four deep fields. For \texttt{HPH}, the $N(z)$ is  either based on the redshift point estimates (median of each PDZ; light red lines) or based on the stack of the individual PDZs (dark red line). For \texttt{SLP}, the $N(z)$ is obtained with the redshift point estimate (blue lines). A comparison is made with COSMOS2020 \texttt{Farmer}/\texttt{EAzY} photo-$z$s (black lines) in the top panels. The COSMOS2020 curves are weighted to match the CLAUDS-HSC ones. The distributions are constructed in bins of $\delta z=0.05$.
    }
    \label{fig:nz}
\end{figure*}
 In Fig.~\ref{fig:nz} we show the galaxy redshift distributions ($N(z)$) for the HSC-CLAUDS catalogs with three different magnitude limits (${m}_U \leq 26.5$, ${m}_i \leq 25.5$, and ${m}_{Ks} \leq 24.5$). The distributions are obtained with the photometric redshifts based on the optical bands only and the optical+NIR bands and with the two configurations (\texttt{HPH} and \texttt{SLP}).
 In addition to the point estimate photo-$z$s, for the \texttt{HPH} configuration we also include the redshift distributions by summing up the individual PDZs. The figure also shows the distribution obtained with \texttt{Farmer}/\texttt{EAzY} photo-$z$s and the same cuts in the photometry. The COSMOS2020 results are weighted such as the integral of the distributions match the average total number of sources between \texttt{HPH} and \texttt{SLP} results.

 The different estimates lead to reasonably consistent redshift distributions, and are in agreement  with the COSMOS2020 redshift distributions. The selection band impacts the overall shape of the distributions, with a bounded $N(z)$ below $z\sim 2.5-3$ with the $u$ band selection and an increasing population at high redshift when adopting a redder selection band.

 We also observe some specific differences. 
 With the NIR selection, a bump at $z\sim3.5$ is visible for the point estimates in the HPH configuration. This bump might be due to poorly informative PDZs (nearly flat) skewed toward high redshift because of the priors, putting the median point estimate of the PDZ in this redshift range. This effect vanishes when adopting the sum of the PDZs.
 
 With the $U$-band selection, we observe a significant difference in the  shape of the $N(z)$ between the photo-$z$s based on the optical bands and those using all bands, the former showing a higher fraction of sources between $1.3\le z\le 2$. 
 This reflects the biased population seen in Fig~\ref{fig:photo-z-bin-mag-i} ($Ugrizy$ plots) and discussed in Sect.~\ref{sec:photoz-comparison-specz}.
 It means that these features are due to the lack of constraints in the absence of NIR data. However, the use of the stacked PDZ helps to reduce this effect.
 
 Finally, we also observe differences between the \texttt{SLP} and \texttt{HPH} redshift distributions in the case of the all-band based photo-$z$s. At low redshift, the \texttt{SLP} distributions increase more rapidly while at high redshift they tend to decrease faster. While the \texttt{Le Phare} code does not apply the \citet{Benitez2000} prior on the redshift likelihood when the NIR bands are included in the computation, \texttt{Phosphoros} always uses the volume prior which penalizes low redshift solutions and favors higher redshift ones. This is the same effect we observe in Fig~\ref{fig:Photo-z-cosmos2020} for the faintest sources. 
%


\section{Data access / catalogs}
\label{sec:data_access}

Two catalogs are presented in this work. They correspond to two different processing methods applied to the same data, providing overall similar data products of equivalent quality as demonstrated in the previous sections. However, the two catalogs also present  differences in the data products (e.g., types of photometry, PDZs, or classification) which can be suited for different purposes. We discuss here these differences that can lead to the selection of a particular catalog for different science cases, though, we recommend the use of both catalogs.

The \texttt{SLP} catalog is extremely similar to the \texttt{Classic}/\texttt{Le Phare} version of the COSMOS2020 catalogs in its production. Therefore, for applications where a close comparison with COSMOS2020 results is desired, this catalog is the most suited. This catalog will also benefit from source physical parameters computed in a similar way as for the COSMOS2020 catalogs (Picouet et al. in prep).

The \texttt{HPH} catalog provides different types of photometry, not present in the \texttt{SLP} one. For instance, it includes cmodel photometry and PSF photometry, that take into account PSF variation across the fields. On the photo-z side, it includes individual PDZ for each object in the catalog. Another interesting feature it the similarities between the HSC pipeline and the LSST pipeline. With its depth and wavelength coverage, the CLAUDS+HSC combination is close to the expected LSST ten-year dataset \citep{Ivezic2019}. Thus, the \texttt{HPH} catalog is most similar to the future LSST data.

The catalogs presented in this work will be publicly released and available for download on several sites.\footnote{
CLAUDS website: \url{https://clauds.net}; \\
DEEPDIP website: \url{https://deepdip.iap.fr} }
For the \texttt{SExtractor} photometry, several catalogs are provided with photometric redshifts estimated with \texttt{Le Phare} 
run with the 6 optical bands and all-bands configurations.

For the \texttt{hscPipe} photometry, all the catalogs are provided with the photometric redshifts estimated with the \texttt{Phosphoros} code. When run with the \texttt{Phosphoros} code, all the catalogs have their full PDZ available in separate files. 

The full description of the columns for the \texttt{SLP} and \texttt{HPH} configurations are given in Appendix~\ref{sec:desccat} together with some practical information not detailed in the core of the paper. 

This release is expected to evolve over time and other catalogs will be available. Additional information such as physical parameters derived with the \texttt{Le Phare} code (Picouet et al, in prep.) and photometric redshifts based on convolutional neural network \citep[][ Ait-Ouahmed et al, in prep.]{Pasquet2019} will also be delivered.

\section{Summary}

In this work, we presented the first public data release of the combined data over the $20~{\rm deg^2}$ of the CLAUDS and HSC surveys. For the four deep and ultraDeep fields of the HSC survey, we processed the CLAUDS data along with the HSC data. For the E-COSMOS and XMM-LSS fields, we also added NIR data parts of the UltraVISTA and VIDEO surveys to extend the wavelength coverage of our catalogs, especially in the ultraDeep regions.

We produced catalogs containing the photometry extracted in two ways. One uses the well-known \texttt{SExtractor} code. The other uses a modified version of the \texttt{hscPipe} that allows the handling of the non-HSC data. CLAUDS and NIR data are pre-processed to match the HSC data format, then are treated by both codes to detect sources and extract their photometry across all bands. We then test the obtained photometry against one another and compare it to the COSMOS2020 photometry. Overall, we find that the photometry from the different methods is in good agreement down to $\rm mag\sim26$ for $Ugrizy$ bands, and down to $\rm mag\sim24.5$ for the NIR ones. For fainter sources, the agreement declines, with increased bias and scatter for the photometry of both codes. For the \texttt{hscPipe} photometry, we found that the obtained photometry in the $grizy$ bands is extremely consistent with that of HSC PDR2. However, the comparisons with our \texttt{SExtractor} photometry and the COSMOS2020 \texttt{Farmer} photometry show some differences probably related to 
the background determination and subtraction. \texttt{hscPipe} seems to systematically over-subtract the background, leading to flux loss for all the sources. We also found good agreement when comparing our galaxy number counts with those from \citet{Weaver2021} \texttt{Farmer} catalog. The strongest differences are found in the $U$ and NIR bands, which are mainly due to the difference in detection strategy and error computation.

We computed photo-$z$s from both our photometry catalogs. Two template-fitting codes were used: \texttt{Phosphoros} was used to produce photo-$z$s from both \texttt{hscPipe} and \texttt{SExtractor} photometry, and \texttt{Le Phare} was used to process \texttt{SExtractor} photometry. The comparison of the different results shows that the major factor for the differences found between the photo-$z$ catalogs is the photometry, which is similar to the observation made by \citet{Weaver2021} with their four different versions of COSMOS2020 photo-$z$s. Here, we found that the photo-$z$s from \texttt{hscPipe} and \texttt{SExtractor} photometry are in good agreement. Whatever the reference redshift they are compared to, we find a scatter $\sigma \lesssim 0.05$ down to ${m}_i\sim25$ or ${m}_{Ks}\sim24$. However, we find that photo-$z$s derived from \texttt{hscPipe} photometry present a systematically higher scatter, of around 0.01, probably due to the difference in the photometry, as both codes perform very similarly on the \texttt{SExtractor} photometry. Another difference in the results, this time attributed to the photo-$z$ code configurations, is the systematic lack of low photo-$z$ solutions from \texttt{Phosphoros} for faint galaxies, due to the priors used in its configuration, which for badly constrained PDZ is favoring higher redshift solutions. All the photo-$z$ comparisons made show that \texttt{Phosphoros} is as efficient as \texttt{Le Phare} and ready to be used as a stand-alone photo-$z$ code. %

The catalogs presented in this work have been publicly released. For the \texttt{hscPipe} photometry, all the data from the images to the fluxes can be fully retrieved in the HSC data access portal. Otherwise, all the CLAUDS images and data product catalogs are accessible on the CLAUDS and DEEPDIP websites and will be updated with upcoming works on photometric redshift and physical parameter estimates. 

%

\begin{acknowledgements}
We thank the anonymous referee for providing helpful comments and suggestions that improved the quality of this work. 
GD acknowledges the support from the Sinergia program of the Swiss National Science Foundation.
This work was supported by the Spin(e) ANR project (ANR-13-BS05-0005), the DEEPDIP ANR project (ANR-19-CE31-0023), and by the Programme National Cosmology et Galaxies (PNCG) of CNRS/INSU with INP and IN2P3, co-funded by CEA and CNES.
MS acknowledges funding support from the Natural Sciences and Engineering Research Council (NSERC) of Canada Discovery Grant and Discovery Accelerator programs, and from the Canada Research Chairs program.   C.C. is supported by the National Natural Science Foundation of China, No. 11803044, 11933003, 12173045, and in part by the National Key R\&D Program of China grant 2017YFA0402704. C.C. acknowledges the science research grants from the China Manned Space Project with NO. CMS-CSST-2021-A05. This work is sponsored (in part) by the Chinese Academy of Sciences (CAS), through a grant to the CAS South America Center for Astronomy (CASSACA).

The Cosmic Dawn Center is funded by the Danish National Research Foundation under Grant No.\,140. JRW and ST acknowledge support from the European Research Council (ERC) Consolidator Grant funding scheme (project ConTExt, grant No. 648179). 

These data were obtained and processed as part of the CFHT Large Area U-band Deep Survey (CLAUDS), which is a collaboration between astronomers from Canada, France, and China described in \citet{Sawicki2019}.  CLAUDS is based on observations obtained with MegaPrime/ MegaCam, a joint project of CFHT and CEA/DAPNIA, at the CFHT which is operated by the National Research Council (NRC) of Canada, the Institut National des Science de l’Univers of the Centre National de la Recherche Scientifique (CNRS) of France, and the University of Hawaii. CLAUDS uses data obtained in part through the Telescope Access Program (TAP), which has been funded by the National Astronomical Observatories, Chinese Academy of Sciences, and the Special Fund for Astronomy from the Ministry of Finance of China. CLAUDS uses data products from CALET and the Canadian Astronomy Data Centre (CADC) and was processed using resources from Compute Canada and Canadian Advanced Network For Astro- physical Research (CANFAR) and the CANDIDE cluster at IAP maintained by Stephane Rouberol.

The Hyper Suprime-Cam (HSC) collaboration includes the astronomical communities of Japan and Taiwan, and Princeton University. The HSC instrumentation and software were developed by the National Astronomical Observatory of Japan (NAOJ), the Kavli Institute for the Physics and Mathematics of the Universe (Kavli IPMU), the University of Tokyo, the High Energy Accelerator Research Organization (KEK), the Academia Sinica Institute for Astronomy and Astrophysics in Taiwan (ASIAA), and Princeton University. Funding was contributed by the FIRST program from Japanese Cabinet Office, the Ministry of Education, Culture, Sports, Science and Technology (MEXT), the Japan Society for the Promotion of Science (JSPS), Japan Science and Technology Agency (JST), the Toray Science Foundation, NAOJ, Kavli IPMU, KEK, ASIAA, and Princeton University. 

This paper makes use of software developed for the Large Synoptic Survey Telescope. We thank the LSST Project for making their code available as free software at  http://dm.lsst.org.

The Pan-STARRS1 Surveys (PS1) have been made possible through contributions of the Institute for Astronomy, the University of Hawaii, the Pan-STARRS Project Office, the Max-Planck Society and its participating institutes, the Max Planck Institute for Astronomy, Heidelberg and the Max Planck Institute for Extraterrestrial Physics, Garching, The Johns Hopkins University, Durham University, the University of Edinburgh, Queen’s University Belfast, the Harvard-Smithsonian Center for Astrophysics, the Las Cumbres Observatory Global Telescope Network Incorporated, the National Central University of Taiwan, the Space Telescope Science Institute, the National Aeronautics and Space Administration under Grant No. NNX08AR22G issued through the Planetary Science Division of the NASA Science Mission Directorate, the National Science Foundation under Grant No. AST-1238877, the University of Maryland, and Eotvos Lorand University (ELTE), and the Los Alamos National Laboratory.
This work is based on data products from observations made with ESO Telescopes at the La Silla Paranal Observatory under ESO program ID 179.A-2005 and on data products produced by CALET and the Cambridge Astronomy Survey Unit on behalf of the UltraVISTA consortium.
Based on data products from observations made with ESO Telescopes at the La Silla Paranal Observatory as part   of the VISTA Deep Extragalactic Observations (VIDEO) survey, under program ID 179.A-2006 (PI: Jarvis)
\end{acknowledgements}

%
%

\bibliographystyle{aa} 
\bibliography{references} 

\begin{appendix}

\section{HSC-SSP PDR2 photometry}
\label{sec:pdr2}
We compare here the photometry we obtain with the \texttt{hscPipe} on the HSC bands to the HSC PDR2 photometry. The comparison is made in the ultraDeep region of the E-COSMOS field. This field is one of the deepest in the $U$-band and is the only field observed across all bands. Thus it could be the most sensitive to any problem caused by considering the extra bands in the \texttt{hscPipe} source detection phase and the forced extraction.

\begin{figure}[]
    \centering
    \includegraphics[width=\linewidth]{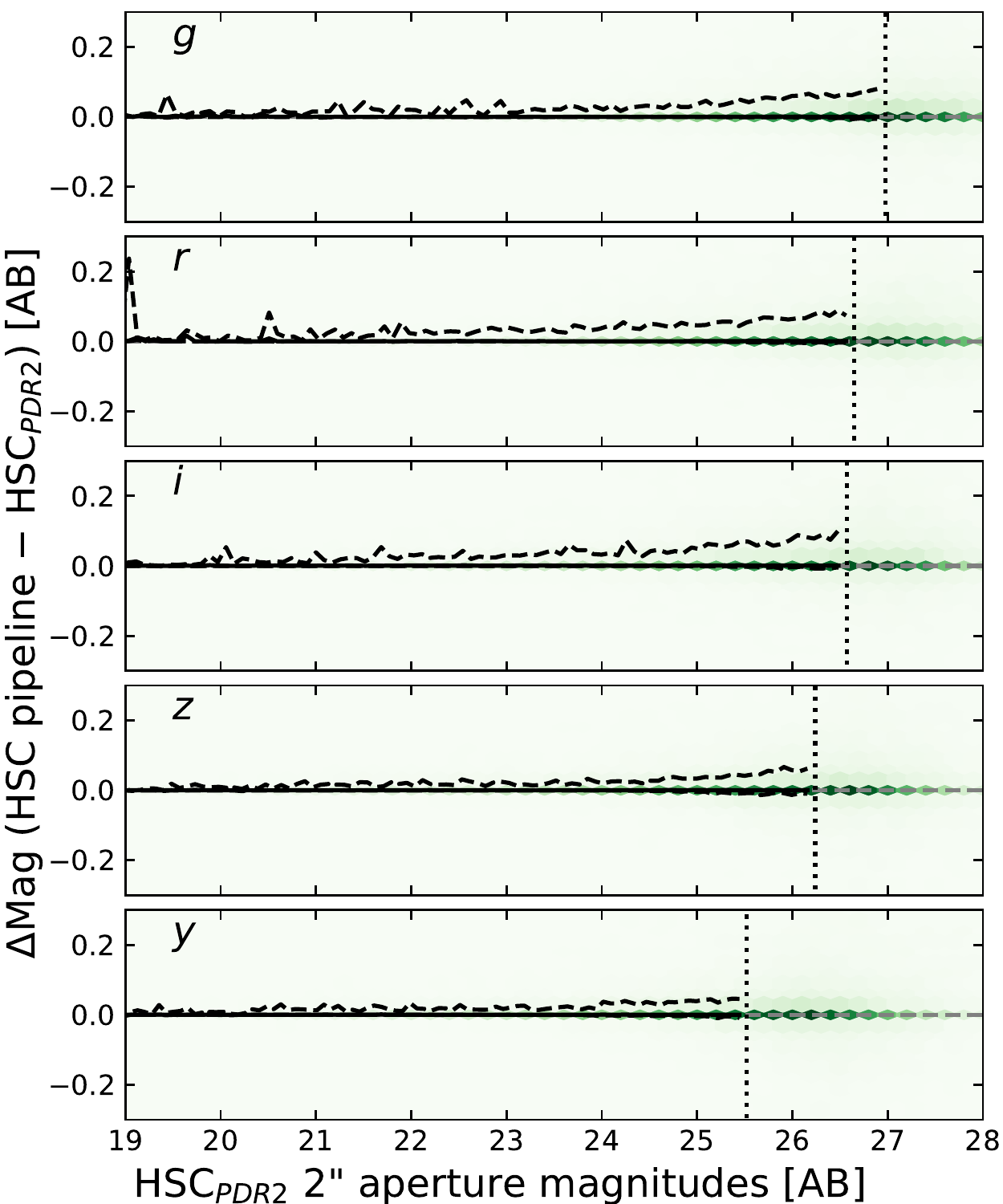}
    \caption{Comparison between our \texttt{hscPipe} photometry and that of HSC PDR2. The dashed lines encompass 68\% of the distributions.}
    \label{fig:hscpipe_vs_psdr2}
\end{figure}

Figure~\ref{fig:hscpipe_vs_psdr2} shows the comparison of the magnitudes in the $grizy$ bands. The distributions show nearly no deviations and an extremely tight scatter between our photometry and the HSC PDR2 one, down to the faintest magnitudes.
As the data from which the photometry is extracted are exactly the same in both cases, the few differences can only be due to the addition of the CLAUDS and NIR bands in the source detection process. The effect can be the new detection of close-by sources or deblending of new sources, resulting in a slightly different allocation of the flux in the two versions of the photometry. 
%
\section{COSMOS2020 \texttt{SExtractor} photometry}
\label{sec:cosmos2020Classic}
\begin{figure}
    \centering
    \includegraphics[width=\linewidth]{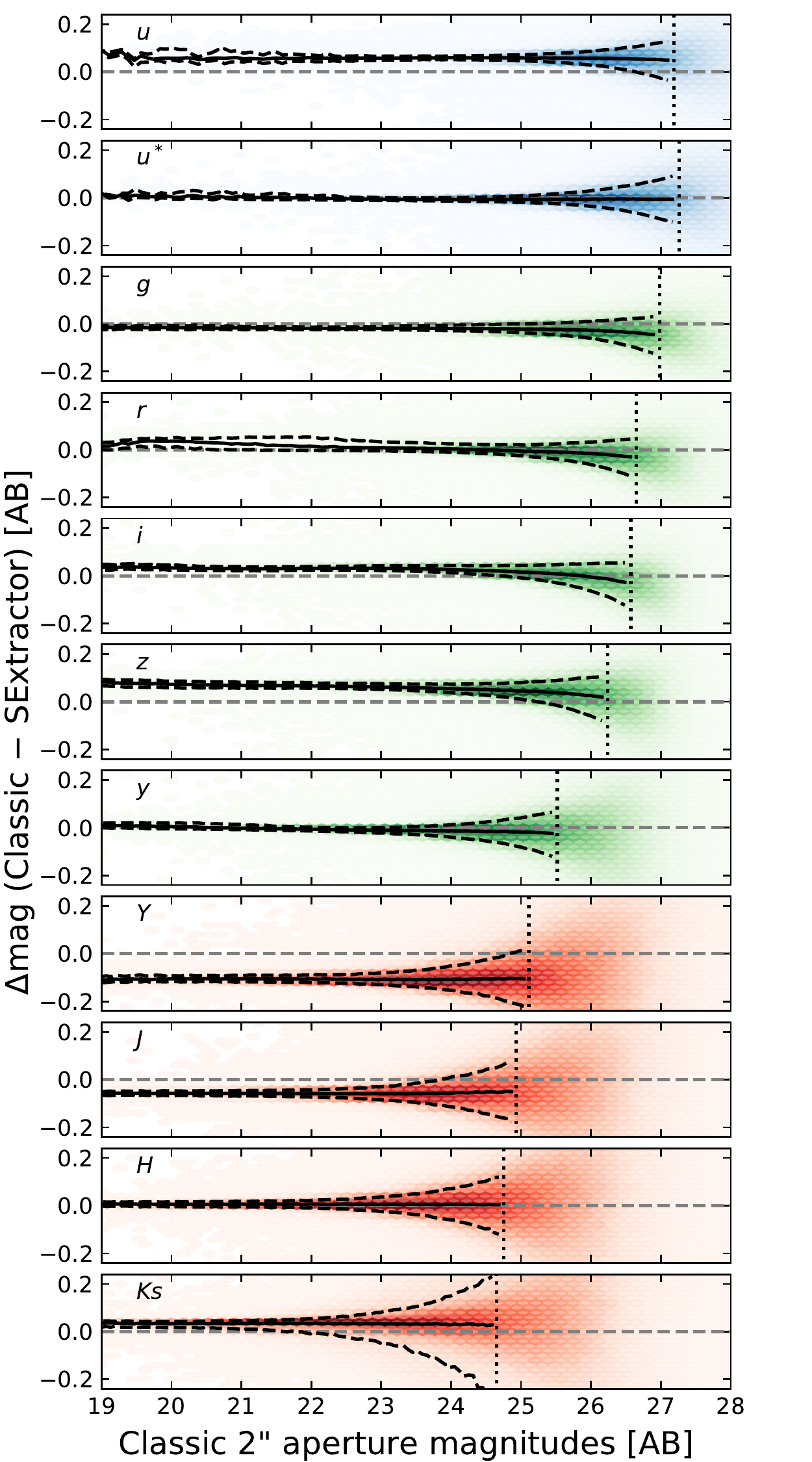}
    \caption{Comparison of COSMOS2020 \texttt{Classic} and \texttt{SExtractor} 2\arcsec\ aperture magnitudes.}
    \label{fig:ClassicSextractor}
\end{figure}

The \texttt{Classic} configuration of COSMOS2020 photometry extraction is extremely similar to our \texttt{SExtractor} configuration. Indeed, the same code is applied to the same data. However, there are a few differences in the settings of the two extractions, for instance, the bands on which the detection is made. \texttt{SExtracor} uses mostly the bluest bands, $Ugrizy$ (+$K_s$ when available), for the detection, whereas \texttt{Classic} uses the reddest bands, $izYJHK_s$, to build its detection map. Another difference is the step of PSF homogenization of the images that is made with the COSMOS2020 \texttt{Classic} methods but not our \texttt{SExtractor} one.

A comparison of the obtained photometry in both cases, like the one presented in Fig.~\ref{fig:ClassicSextractor}, shows the differences induced by these configuration choices. The figure presents the comparison of the 2\arcsec\ photometry between \texttt{Classic} and \texttt{SExtractor}. As expected, the relation between the two photometry catalogs is tight. However, we can observe strong biases, as large as $\sim0.07$mag (e.g., for the $z$ or the $Y$ bands). This effect is due to PSF homogenization. 
%
\section{SExtractor-\texttt{hscPipe} 2 \arcsec\ photometry}
\label{sec:comparison2arcsec}
\begin{figure}
    \centering
    \includegraphics[width=\linewidth]{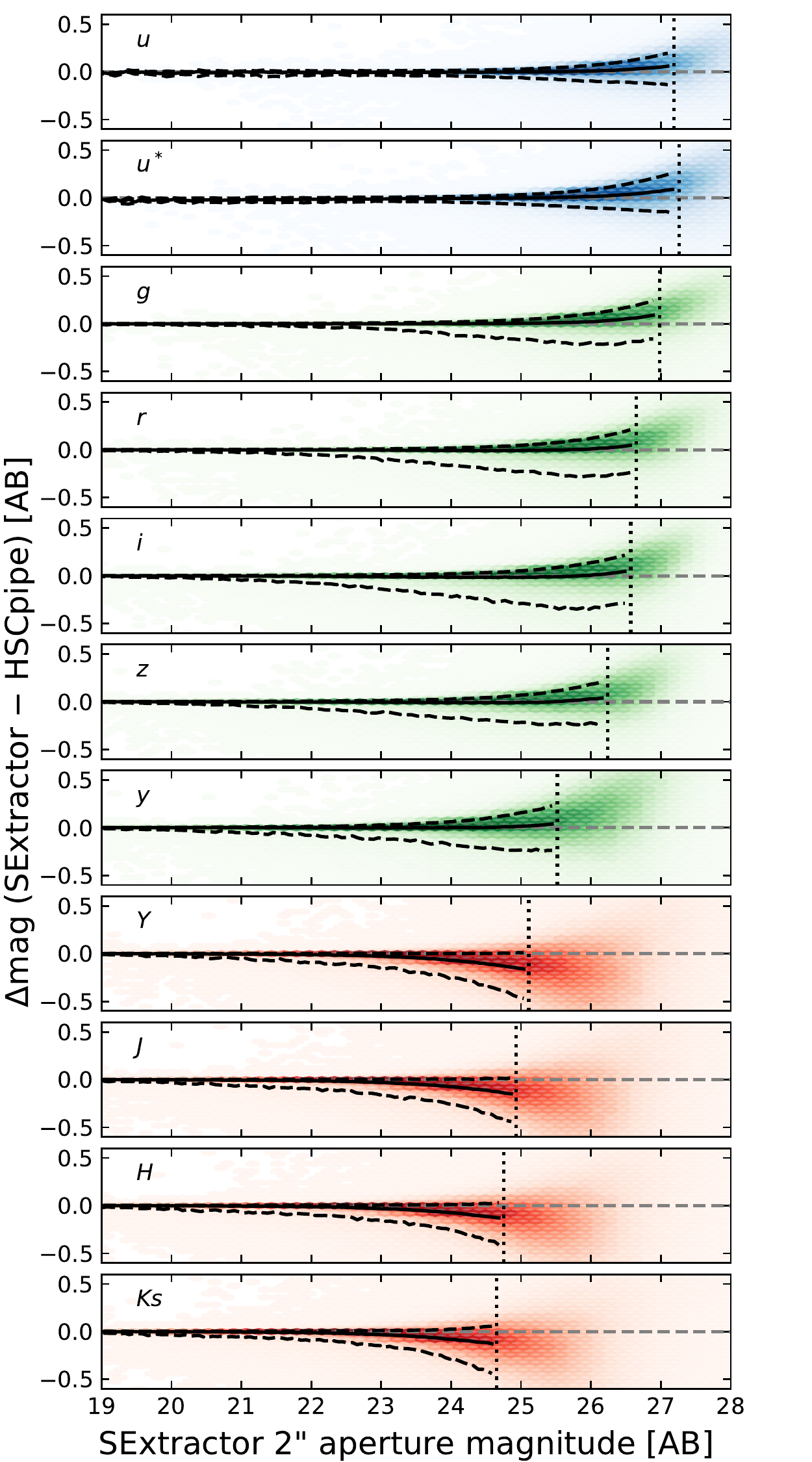}
    \caption{Comparison of \texttt{SExtractor} and the \texttt{hscPipe} 2\arcsec\ aperture magnitudes.}
    \label{fig:compSEx_HSC_2arcsec}
\end{figure}
The 2\arcsec\ aperture photometry is the one used in both configurations to compute photometric redshifts. In Fig.~\ref{fig:compSEx_HSC_2arcsec}, we present the direct comparison of the 2\arcsec\ aperture photometry from \texttt{SExtractor} and \texttt{hscPipe}. We note a tight relation between the measurements obtained with these two codes; however, an asymmetric scatter is present. This asymmetry seems to indicate that the \texttt{hscPipe} fluxes are lower than the \texttt{SExtractor} ones, and this effect is stronger for fainter sources. As these sources are more sensitive to background noise, we interpret the increase in asymmetry toward fainter magnitudes as a sign of stronger subtraction of the background by \texttt{hscPipe} compared to \texttt{SExtractor}. The same conclusion was reached when analyzing Fig.~\ref{fig:comparisonFarmerHsc},~\ref{fig:comparisonSexHsc} and~\ref{fig:comparionColors} in Sect.~\ref{sec:photometrty-quality}.
%
\section{\texttt{Phosphoros} results on \texttt{SExtractor} photometry}
\label{sec:PhosphorosSExtractor}
Throughout the paper, the results that are presented are those of \texttt{Phosphoros}/\texttt{hscPipe} and \texttt{Le Phare}/\texttt{SExtractor}. This choice makes it difficult to attribute the differences observed in the photo-$z$s to the codes or to the photometry. To address this problem, we apply here \texttt{Phosphoros} to the \texttt{SExtractor} photometry to create a version of the photo-$z$s for which only the photo-z codes differ. 
The configuration used to compute these photo-$z$s is the same as that presented in Sect.~\ref{sec:photoz-phosphoros}. We use \texttt{SExtractor} 2\arcsec\ aperture photometry uncorrected for galactic extinction, that we fit with the templates, extinction law, and priors described in Sect.~\ref{sec:photoz-phosphoros}. As a result, PDZs are produced, from which the point estimates (median) are computed.

\begin{figure*}
    \centering
    \includegraphics[width=\linewidth]{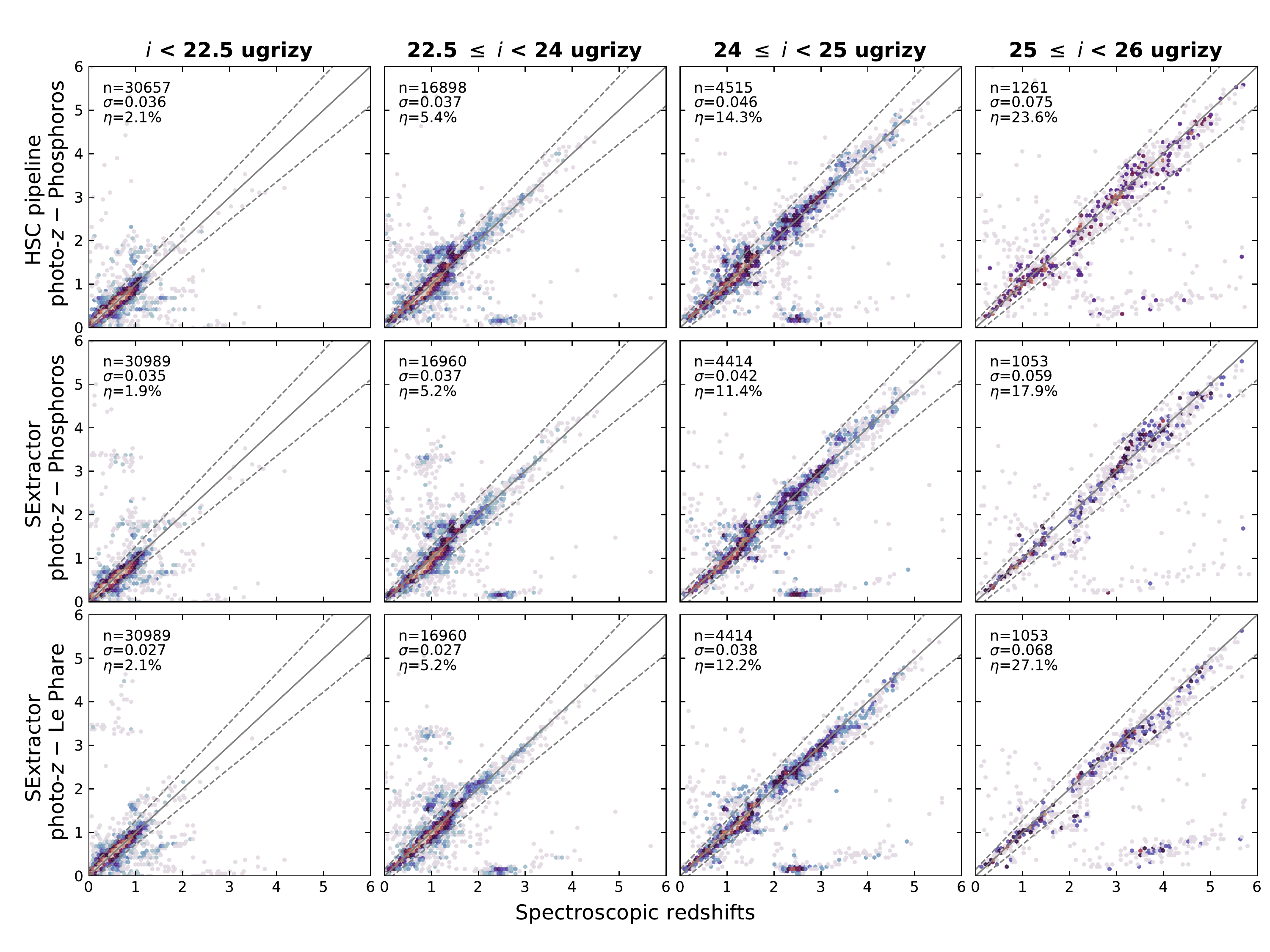}
    \caption{Comparison of \texttt{Phosphoros} results when computing photo-$z$s on \texttt{hscPipe} (Top) and \texttt{SExtractor} (Middle) photometry. The results are also compared to those of \texttt{Le Phare} and \texttt{SExtractor} (Bottom). The comparison is made using photo-$z$s computed using only the $Ugrizy$ bands in all three cases.}
    \label{fig:PhosphorosSExtractor}
\end{figure*}

Figure~\ref{fig:PhosphorosSExtractor} presents a comparison of the results of \texttt{SExtractor}/\texttt{Phosphoros} (middle panels) with both \texttt{hscPipe}/\texttt{Phosphoros} (top panels) and \texttt{SExtractor}/\texttt{Le Phare} (bottom panels). In all three cases, $Ugrizy$ photo-$z$s are compared with the spectroscopic sample. In these plots, we can see that the two sets of photo-$z$s computed from the \texttt{SExtractor} photometry have the same quality when comparing the scatter and the outlier fractions. The scatter from the \texttt{Phosphoros} results is still larger in the bright bins, whereas for the fainter sources the \texttt{Phosphoros} results present a better scatter and outlier fraction than \texttt{Le Phare} photo-$z$s. This is probably due to the choices in the priors applied by both codes.

These results show that the majority of the differences observed between the photo-$z$s presented in this work are due to the differences in photometry. Outside of the effect of the priors, there are a few differences between \texttt{Phosphoros} and \texttt{Le Phare} photo-$z$s, which is expected due to the strong similarities between the two template-fitting codes. All these comparisons thus validate that \texttt{Phosphoros} is able to provide sensible results and is ready to be used by the community.
%
\section{Description of the released catalogs}
\label{sec:desccat}
The current release (Sect~\ref{sec:data_access}) consists of a set of catalogs for the four HSC-CLAUDS regions. Several versions are available depending on the adopted photometry (based on the \texttt{HSC pipeline} (Sect.~\ref{sec:photometry-hsc}) or \texttt{SExtractor} run  (Sect.~\ref{sec:photometry-sextractor})), the photometric code used ( \texttt{Phosphoros} (Sect.~\ref{sec:photoz-phosphoros}) or \texttt{Le Phare}  (Sect~.\ref{sec:photoz-lephare})) and the number of photometric bands considered for the photometric redshift runs ($Ugrizy$ and all-bands). Only sub-regions in the XMM-LSS and E-COSMOS fields are covered with the near-infrared VISTA imaging.  

 The description of the catalog outputs is given in Tab~\ref{tab:catalogSExtractor} for the \texttt{SExtractor} -\texttt{Le Phare} catalogs and in Tab~\ref{tab:catalogHSCpipe} for the \texttt{hscPipe} -\texttt{Phosphoros} catalogs\footnote{the catalogs do not contain all the information produced by the \texttt{hscPipe} output, but the raw photometry file are available through the HSC collaboration distribution portal.\url{https://hsc-release.mtk.nao.ac.jp/doc/index.php/data-access__pdr3/} }.
 
 As the footprints of the different surveys are different, the keyword 
 \textsc{FLAG\_FIELD\_BINARY} allows the user to restrict the data to a specific dataset by combining its 7 boolean elements (described in Tab~\ref{tab:catalogSExtractor}). A version defined with a single integer is also available (\textsc{FLAG\_FIELD} for \texttt{Le Phare} catalogs). 
 To guarantee the selection of clean objects, the flags \textsc{MASK} and \textsc{ST\_TRAIL} must be set to 0 (for \texttt{SExtractor} catalogs) or \textsc{isOutsideMask} to "True" (for \texttt{hscPipe} catalogs).
 
All the catalogs include a stellar classification based on $\chi^2$ estimates between the Pickles stellar templates \citep{Pickles1998} and the galaxy templates. If the sources are compact and better fitted by a stellar template they are flagged as stars (\textsc{OBJ\_TYPE}=2 or \textsc{isStar}=True).
In addition the \texttt{Le Phare} code also performs a classification using a QSO library \citep{Salvato2011}. If the sources are compact and better fitted by a QSO template, they are flagged as QSO (\textsc{OBJ\_TYPE}=1). 
  
For the catalogs based on the \texttt{Phosphoros} code, the PDZs of all sources are provided in separate catalogs for each tract of the HSC data system. They only contain the source IDs and the PDZ values within the redshift range $0\le z \le 6$ in steps of $\delta z=0.01$.
For the catalogs based on the \texttt{Le Phare} code, we also provide the measurement of the physical parameters (Picouet et al., in prep.) derived with synthetic SEDs from the \citet[]{Bruzual2003} library.

%

\begin{table*}
\small
  \caption{Summary of the \texttt{SExtractor}/\texttt{Le Phare} catalog.}
\begin{tabular}{ l l l}
\textbf{Column name} & \textbf{Unit} & \textbf{Description}\\
\hline
\multicolumn{3}{c}{\textbf{General Parameters}}\\ 
\textsc{ID} & - & Identifiant \\
\textsc{RA, DEC}& deg & Right ascension and declination of barycenter (J2000) \\
\textsc{TRACT, PATCH} & - & Position in the HSC frames\\
\textsc{MASK} & [0,1] & Bright stars mask, clean area: MASK==0\\
\textsc{FLAG\_FIELD\_BINARY} & Bool[x7] & [HSC, u, uS, J, VIRCAM, u$^*_{\rm deep}$, HSC$_{\rm deep}$]\\
\textsc{FLAG\_FIELD} & integer & $\Sigma_{{\rm HSC}, u, u^*, J, {\rm VIRCAM}, u^*_{\rm deep}, {\rm HSC}_{\rm deep}} 2^{FLAG(band)}$ \\
\textsc{Z\_SPEC} & - & Spectroscopic redshift for matched objects\\

\\
\multicolumn{3}{c}{\textbf{Photometry Parameters [SExtractor]}}\\ 
\textsc{A\_WORLD, B\_WORLD} & deg & Profile RMS along ellipse's axis \\
\textsc{KRON\_RADIUS} & deg & Kron apertures in units of A or B\\
\textsc{THETA\_WORLD} & deg & Ellipse orientation\\
\textsc{ELONGATION} & - & \texttt{A\_WORLD/B\_WORLD}\\
\textsc{ELLIPTICITY} & - & \texttt{1 - B\_WORLD/A\_WORLD}\\
\textsc{EB\_V} & - & Galactic extinction (Schlegel et al. 1998)\\
\textsc{FWHM\_WORLD\_HSC\_I} & deg & FWHM assuming a gaussian core\\
\textsc{FLUX\_RADIUS\_[0.25,0.5,0.75]\_HSC\_I} & pix & Fraction-of-light radii\\
\textsc{CLASS\_STAR\_HSC\_I} & pix & \texttt{SExtractor} star estimator\\
\textsc{[band], [band]\_err} & mag & total mag, galactic extinction corrected\\
\textsc{MAG\_APER\_2s\_[band], MAGERR\_APER\_2s\_[band]} & mag & magnitude and magnitude error measured in a 2'' aperture\\
\textsc{MAG\_APER\_3s\_[band], MAGERR\_APER\_3s\_[band]} & mag & magnitude and magnitude error measured in a 3'' aperture\\
\textsc{MAG\_ISO\_[band], MAGERR\_ISO\_[band]} & mag & magnitude and magnitude error measured in isophotal aperture\\
\textsc{Offset\_[2S,3S,ISO]} & mag & Total magnitude offsets to be added to the respective magnitudes\\

\\
\multicolumn{3}{c}{\textbf{Parameters computed with \texttt{LePhare} - Zphot}}\\ 

\textsc{Z\_BEST, Z\_BEST68\_LOW, Z\_BEST68\_HIGH} & - & Maximum of the ML distribution \\
\textsc{NBAND\_USED} & integer & Number of bands\\
\textsc{CHI\_BEST, CHI\_STAR, CHI\_QSO} & - & Chi2 of the best fit with galaxy, star, and QSO template\\
\textsc{MOD\_BEST, MOD\_STAR, MOD\_QSO} & - & Best fit template\\
\textsc{Z\_ML, Z\_ML68\_LOW, Z\_ML68\_HIGH} & - & Median of the ML distribution \\
\textsc{Z\_SEC} & - & Redshift of the secondary peak\\
\textsc{Z\_QSO} & - & QSO redshift computed\\
\textsc{ZPHOT} & - & Combination of photometric redshift: $z_{ML}\rightarrow z_{BEST}$ \\

\\
\multicolumn{3}{c}{\textbf{Parameters computed with \textsc{LePhare} - BC03}}\\ 
\textsc{Z\_BC03} & - & best z combination used for BC03 run: $z_{SPEC}\rightarrow z_{PHOT}$ \\
\textsc{MAG\_ABS\_[band]} & mag & \texttt{LePhare} computed absolute magnitude\\
\textsc{MOD\_BEST\_BC03} & integer & Best fit template\\
\textsc{EBV\_BEST} & - & Best fit template\\
\textsc{EXTLAW\_BEST} & -& Best fit template\\
\textsc{AGE\_X} & ${\rm year}$ & Age of best fit with X\ in (BEST, MED, INF, SUP)\\
\textsc{MASS\_X} & ${\rm M_\odot}$ & Mass of best fit with X\ in (BEST, MED, INF, SUP)\\
\textsc{SFR\_X} & ${\rm M_\odot\,year^{-1}}$ & SFR of best fit with X\ in (BEST, MED, INF, SUP)\\
\textsc{SSFR\_X} & ${\rm year^{-1}}$ & sSFR of best fit with X\ in (BEST, MED, INF, SUP)\\
\textsc{LUM\_NUV\_BEST} & - & NUV luminosity\\
\textsc{LUM\_R\_BEST} & - & R luminosity\\
\textsc{LUM\_K\_BEST} & - & K luminosity\\

\multicolumn{3}{c}{\textbf{Other parameters}}\\ 

\textsc{OBJ\_TYPE} & [0,1,2] & Object flag: 0 for galaxies, 1 for QSOs, 2 for stars\\
\textsc{COMPACT} & [0,1] & Flag for compact object (see eq 1)\\
\textsc{PASSIVE} & [0,1] & Flag for passive VS star-forming galaxies based on NUVrK\\
\hline
  \end{tabular}
\label{tab:catalogSExtractor}
\end{table*}

\begin{table*}
\small
  \caption{Summary of the \texttt{hscPipe}/\texttt{Phosphoros} catalog. }
\begin{tabular}{ l l l}
\hline
\hline
\rule{0pt}{1.2em}\textbf{Column name} & \textbf{Unit} & \textbf{Description}\\
\hline
 \multicolumn{3}{c}{\rule{0pt}{1.2em}\textbf{Id and locations}}\\ 
\rule{0pt}{1.2em}\textsc{ID} & --- & Identifiant \\
\textsc{RA, DEC}& deg & Right ascension and declination of centroid (J2000) \\
\textsc{tract, patch} & --- & Tract and patch ids in the \texttt{hscPipe} system\\

\\
\multicolumn{3}{c}{\textbf{Photometry}}\\ 
\rule{0pt}{1.2em}\textsc{FLUX\_APER\_2\_[band], FLUXERR\_APER\_2\_[band]} & $\mu {\rm Jy}$ & Flux and flux error measured in a 2\arcsec\ aperture\\
\textsc{FLUX\_APER\_3\_[band], FLUXERR\_APER\_3\_[band]} & $\mu {\rm Jy}$ & Flux and flux error measured in a 3\arcsec\ aperture\\
\textsc{FLUX\_PSF\_[band], FLUXERR\_PSF\_[band]} & $\mu {\rm Jy}$ & Flux and flux error measured using PSF modeling\\
\textsc{FLUX\_KRON\_[band], FLUXERR\_KRON\_[band]} & $\mu {\rm Jy}$ & Flux and flux error measured in Kron apertures\\
\textsc{RADIUS\_KRON\_[band]} & \arcsec\ & Kron radius\\
\textsc{FLUX\_CMODEL\_[band], FLUXERR\_CMODEL\_[band]} & $\mu {\rm Jy}$ & Flux and flux error measured using cmodel fitting\\
\textsc{hasBadPhotometry\_[band]} & Bool & Flag indicating problem in photometry measurement\\
\textsc{isDuplicated\_[band]} & Bool & Flag indicating if source is duplication of primary source\\
\textsc{isNoData\_[band]} & Bool & Flag indicating if source is not extracted from observed data\\
\textsc{isSky\_[band]} & Bool & Flag indicating if source is extracted from sky\\
\textsc{isParent\_[band]} & Bool & Flag indicating if source a parent blended source\\
\textsc{notObserved\_[band]} & Bool & Flag indicating if source is observed in the band\\
\textsc{isClean\_[band]} & Bool & Flag indicating if all photometry flags are false\\

\textsc{isCompact\_[HSC-band]} & Bool & Flag indicating if source is compact in HSC band\\
\textsc{isCompact} & Bool & Flag indicating if source is compact all HSC bands\\
\\
\multicolumn{3}{c}{\textbf{Results computed with Phosphoros}}\\ 

\rule{0pt}{1.2em}\textsc{ZPHOT\_NIR, Z\_LOW68\_NIR, Z\_HIGH68\_NIR}$^a$ & --- & PDZ median and limits encompassing 68\% of PDZ using all bands\\
\textsc{Z\_CHI\_NIR}$^a$ & --- & Point estimate with highest likelihood using all bands\\
\textsc{Z\_PEAK\_NIR}$^a$ & --- & Point estimate with highest PDZ value using all bands\\
\textsc{Posterior-Log\_NIR}$^a$ & --- & Log of best posterior value \\
\textsc{Likelihood-Log\_NIR}$^a$ & --- & Log of best likelihood value\\

\rule{0pt}{1.2em}\textsc{ZPHOT\_6B, Z\_LOW68\_6B, Z\_HIGH68\_6B}$^a$ & --- & PDZ median and limits encompassing 68\% of PDZ using $Ugrizy$ bands\\
\textsc{Z\_CHI\_6B}$^a$ & --- & Point estimate with highest likelihood using $Ugrizy$ bands\\
\textsc{Z\_PEAK\_6B}$^a$ & --- & Point estimate with highest PDZ value using $Ugrizy$  bands\\
\textsc{Posterior-Log\_6B}$^a$ & --- & Log of best posterior value using $Ugrizy$ bands\\
\textsc{Likelihood-Log\_6B}$^a$ & --- & Log of best likelihood value using $Ugrizy$ bands\\

\rule{0pt}{1.2em}\textsc{ZPHOT, Z\_LOW68, Z\_HIGH68} & --- & PDZ median and limits encompassing 68\% of PDZ using all bands\\
\textsc{Z\_CHI} & --- & Point estimate with highest likelihood using all bands\\
\textsc{Z\_PEAK} & --- & Point estimate with highest PDZ value using all bands\\
\textsc{Posterior-Log} & --- & Log of best posterior value \\
\textsc{Likelihood-Log} & --- & Log of best likelihood value\\

\rule{0pt}{1.2em}\textsc{Likelihood-Log\_star} & --- & Log of best likelihood value star templates with all bands available\\

\\
\multicolumn{3}{c}{\textbf{Other info}}\\ 

\rule{0pt}{1.2em}\textsc{isOutsideMask} & [0,1] & Source outside of updated bright stars mask and satellite trail footprint\\
\textsc{FLAG\_FIELD\_BINARY} & Bool[x7] & [HSC, $u$, $u^*$, $J$, VIRCAM, $u^*_{\rm deep}$, HSC$_{\rm deep}$]\\
\textsc{isStarTemp} & Bool & Have a better Likelihood-Log with star template than galaxy templates\\
\textsc{isStar} & Bool & \textsc{isStarTemp} \& \textsc{isCompact} \\
\hline
\multicolumn{3}{l}{\rule{0pt}{1.2em}$^a$ Only for E-COSMOS and XMM-LSS fields as they both benefit from additional NIR data.}
  \end{tabular}
\label{tab:catalogHSCpipe}
\end{table*}
%
%

\end{appendix}

\end{document}